\documentclass[aps,pre,
numerical, 
floatfix,notitlepage,twocolumn,%
longbibliography
]{revtex4-1}

\usepackage{bm}
\usepackage{graphicx}
\usepackage{color}

\newcommand\beq{\begin{equation}} 
\newcommand\eeq{\end{equation}}  
\newcommand\bea{\begin{eqnarray}}  
\newcommand\eea{\end{eqnarray}}

\begin{document}

\title{Front-propagation in bacterial inter-colony communication}
\author{Vera Bettenworth, Matthew McIntosh, Anke Becker and Bruno Eckhardt}
\affiliation{LOEWE-Zentrum f\"ur Synthetische Mikrobiologie (SYNMIKRO), Philipps-Universit\"at Marburg,
D-35032 Marburg, Germany}
\date{\today}

\begin{abstract}
Many bacterial species exchange signaling molecules to coordinate population-wide responses. For this process known as quorum sensing the concentration of the respective molecules is crucial. Here we consider the interaction between spatially distributed bacterial colonies so that the spreading of the signaling molecules in space 
becomes important. The exponential growth of the signal-producing populations and the corresponding increase in signaling molecule production result in an exponential concentration profile that spreads with uniform speed. The theoretical predictions are supported by experiments with different strains of the soil bacterium \textit{Sinorhizobium meliloti} that display fluorescence when either producing or responding to the signaling molecules.\\[1em]
{\textit{Dedicated to Hans Braun on occasion of his seventieth birthday}}
\end{abstract}

\maketitle

\textbf{In a process called quorum sensing, bacteria exchange signaling molecules 
to collect feedback on the size of their community and to initiate a population-wide change in behavior
once a certain \textit{quorum} has been reached. 
A variety of signaling molecules and different pathways for the production and detection of these molecules have been described for different species, but these studies have also shown
that there are common features underlying many quorum sensing systems.
Here, we focus on general spatiotemporal aspects of this communication, the transmission of information between far-scattered bacterial colonies over large cell-free distances where the main mode of signal propagation is diffusion. As we describe, the exponential growth of the colonies producing the signaling molecules
has a profound effect on the  way the signal spreads in space: While
a constant source results in a distribution where the signaling molecules 
become more and more dilute with increasing distance from the source, the continuous boost in production 
by an exponentially growing colony conspires with diffusion to produce a front
that travels from the source with constant speed. Experiments with 
the model bacterium \textit{Sinorhizobium meliloti} with localized sources and
spatially distributed receiver colonies show a position-dependent response that
is in agreement with the main predictions from the theory. }

\section{Introduction}
\label{intro}
Living organisms employ a large variety of chemical, electrical, optical or mechanical processes to sense their 
abiotic and biotic environment and to respond to this manifold of cues. 
In Hans Brauns work the focus has been on electrical signals, how they are generated by ion channels in 
neuronal membranes, 
and how they vary in response to external stimuli \cite{Braun:1994tz,Neiman:1999bh,Braun:1997tn,Braun:1999ta}.
He and collaborators developed Hodgkin-Huxley-based mathematical models that show rich dynamics
\cite{Braun:2000hc,Feudel:2000hs} and are in sufficiently good agreement with observations that they 
have been implemented in a suite of simulation software for  physiological experiments, 
available at \texttt{http://www.virtual-physiology.com/}. 
However, organisms use such processes 
not only for sensing, but also to communicate and interact - they set 
up chemical or electrical waves to transport signals coherently over large distances. As in the case
of the sensory systems, the response and shape of the signals is a consequence of 
nonlinear excitatory dynamics \cite{Keener:2008xy}. {In this context, it is interesting to note that ion channels and potassium waves have been reported for bacterial biofilms as well, apparently providing a kind of neuron-like electrical signaling \cite{Prindle:2015hp,Liu:2015ht,Humphries:2017gp}.}

The bacterial communication system we explore is of purely chemical nature and generally seen as the means 
for bacteria to determine the size or density of their population, and to regulate their behavior accordingly: Individual bacterial cells produce signaling molecules, so-called autoinducers, that spread in their environment where the molecules can be detected by other bacterial cells. If the concentration of the signaling molecules is high enough \cite{Kaplan:1985ve}, they induce a more or less population-wide change in behavior, e.g., the production of virulence factors in pathogenic bacteria, or of extracellular matrix in the course of biofilm formation. Since the signal concentration, and thus the onset of the response, is related to the number of cells producing the signaling molecules, the term 
`quorum sensing` has been coined for this process 
\cite{Fuqua:1994wb}. 

Quorum sensing systems have turned out to be so ubiquitous that by now they are thought of as 
\textit{not the exception but, rather, the norm in the bacterial world}
\cite{Bassler:2006kd}. 
The chemical nature of the signaling molecules as well as the number of signaling pathways employed differs from 
species to species, but many quorum sensing systems share a common set of elements. For instance, the regulatory 
circuits usually include a positive feedback loop of the signaling molecule on its own production, 
bringing about a rather defined transition into the quorum-sensing state \cite{Miller:2001uo,Long:2009bp}.  Note that while the response of the organisms to the signaling molecules thus has nonlinear elements, the diffusive spreading of the signal in the organisms' environment is of an essentially linear nature, in contrast to the nonlinear signal propagation in neuronal or chemical waves.

Research on the processes involved in quorum sensing has long been reinforced by theoretical studies, 
taking into account factors like flow, diffusion, adhesion, decay or 
degradation of the signaling molecules, or growth of the signal-producing population (reviewed in 
\cite{PerezVelazquez:2016jm}). However, these studies often focus on processes in 
single cells, or cells in well-mixed liquid systems. Only few studies include or specifically investigate 
spatiotemporal signal propagation within single colonies or populations \cite{Basu:2005cq,Danino:2010km,Dilanji:2012ds,Langebrake:2014fg,Ramalho:2016ki}. 
Of particular relevance to our study of inter-colony communication is the description of diffusive 
spreading of signaling molecules beyond individual colonies in \cite{Alberghini:2009gi},
where the spatial variation of the concentration and its dependence on the number of 
cells producing the signal are analyzed.

Diffusion will typically act to dilute the molecules and
hence to attenuate the signal. For instance, 
for a constant source, the distance of a given concentration level from the source 
will increase with the square root of time. 
This is very different from the signals along nerve fibers or in chemical waves,
where the propagation occurs at a constant speed. However, as we will deduce in the subsequent theoretical analysis, the interplay between diffusion and the exponentially increasing activity of a growing bacterial colony will set up a propagating front as well. This conclusion is supported by time-lapse experiments with sender and receiver colonies of the model bacterium \textit{Sinorhizobium meliloti} that display an increase in fluorescence when entering the quorum-sensing state: The spatiotemporal pattern of the response of the receiver colonies to the signaling molecules produced by the sender colonies likewise suggests that the threshold signal concentration necessary to trigger the quorum sensing response spreads in the environment with uniform speed. 
 
The outline of the paper is as follows: In section \ref{model} we discuss the modeling of the diffusion process and describe the spatiotemporal variation of concentrations for different sources, including the formation of fronts by exponentially growing sources. In section \ref{sino} we describe our model system \textit{S. meliloti} and the experimental setup. Results from the observations and the modeling are combined and discussed in section \ref{results}. We conclude with a few general observations in section \ref{conclusion}. Details of the experimental materials and methods are given in section \ref{methods}. 

\section{Modelling diffusive spreading from source colonies}
\label{model}
As stated above, quorum sensing signaling molecules produced and released from bacterial cells spread diffusively in the environment. By Fick's law, the concentration $c(\mathbf{x},t)$ of molecules at 
spatial position $\mathbf{x}$ and time $t$ obeys the diffusion equation
\beq
\partial_t c = D \Delta c + \tilde q(\mathbf{x},t)
\label{diffusion_equation}
\eeq
with diffusion constant $D$ and a term $\tilde q(\mathbf{x},t)$ that contains the temporal
variations and the spatial distribution of one or more sources. 

Many examples of such diffusive processes in different geometries and dimensions are discussed in the classic text of Carslow and Jaeger \cite{Carslow:1959vx}. Since the experimental setup we employ here mainly involves diffusion in the plane, we will subsequently focus on two-dimensional cases.

As a first situation we consider a single signal pulse, in which the concentration of molecules 
that are initially localized will spread with a Gaussian shape. In two dimensions the concentration
is given by
\beq
c(r,t) = \frac{1}{4\pi Dt} e^{-\frac{r^2}{4Dt}}
\eeq
with $r$ the distance from the source.
Levels of constant concentration $c(r,t)=c_0$ are circular, and move in time according to
\beq
r_0^2=4 D t \, \left(\ln (1/c_0)- \ln 4 \pi D t \right)
\eeq
Leaving aside the second term, which is only important for very short and very long times,
we find that $r_0(t) \propto \sqrt{t}$.  Moreover, the gradient at that position and time, which
relates uncertainties in concentration to variations in space according to
$\delta r = \delta c / |\partial c/\partial r|$, is given by
\beq
\left. \frac{\partial c}{\partial r}\right|_{r_0} = - \frac{1}{\sqrt{D t}}
\eeq
and decreases with time, so that the concentration level becomes less well defined.

In contrast to a single signal pulse, a time-dependent source $q(t)$ at the origin contributes 
new molecules at every time step, and their contributions have to be added up: 
if the source is turned on at time $t_0$,
the concentration profile is given by
\beq
c(r,t) = \int_{t_0}^{t} \frac{1}{{4\pi D (t-t')}} e^{-\frac{r^2}{4D(t-t')}} q(t') dt'
\label{diff_2d_soln}
\eeq
For most sources, this has to be evaluated numerically (see below). An analytical solution is
possible for our case where
the sources of the signal are exponentially growing colonies of bacteria that release - if we assume a constant mean production rate once the cells have entered the quorum-sensing state - 
an exponentially growing number of signaling molecules. Such a source is described by an
exponentially growing strength $q(t) = q_0 \exp \lambda t$. If we assume that it has
been active forever ($t_0\rightarrow -\infty$),
the concentration becomes
\beq
c(r,t) = q_0 e^{\lambda t} \int_{0}^{\infty} \frac{1}{{4\pi D \tau}} e^{-\frac{r^2}{4D\tau}} 
e^{-\lambda \tau} d\tau
\eeq
which can be integrated exactly:
\beq
c(r,t) = \frac{2q_0}{4\pi D} e^{\lambda t} \mbox{K}_0(r/\ell) \approx
\frac{2q_0}{4\pi D} \sqrt{\frac{\pi \ell}{2 r}} e^{\lambda t -r/\ell}  
\label{c_int}
\eeq
where $\mbox{K}_0(\xi)$ is the modified Bessel function of index $0$ and $\ell = \sqrt{D/\lambda}$
a characteristic length. Note that space and time separate: There is an overall exponential increase
in time with the same rate $\lambda$ as for the source. The shape of the profile in space for fixed time
is given by the Bessel function, which for large distances has an exponential decay (modulo a weaker
$1/\sqrt{r}$ factor).
Levels of constant concentration are circular, and spread outwards like
$r_0=\sqrt{D \lambda}t$, with a constant speed
\beq
v=\sqrt{D \lambda}\,.
\label{velocity}
\eeq
Moreover, the gradient at the position of the isocontour is essentially constant, 
$\partial c /\partial r\approx 1/\ell$. A constant speed of propagation and a preserved
gradient link this process to front propagation in pattern forming systems which shares
the same properties.

The analysis given here is reminiscent of the front propagation described by 
Kendall \cite{Kendall:1948df}, which has also been discussed in the context 
of quorum sensing \cite{Fu:2012en}:
in a diffusion equation where the signal is amplified everywhere, so that the source
$\tilde q = \lambda c$ is proportional to the concentration with an amplification factor $a$,
he describes that there are propagating solutions $c(x,t) = f (x-vt)$. They exist for a
range of velocities and the one that dominates in the long run has the
smallest velocity, $v_c = 2 \sqrt{D \lambda}$. The front speed has the same dependence
on diffusion and growth rate as (\ref{velocity}) but it is twice as large, which is due to 
the difference in the amplification process: In the Kendall problem, the concentration
is amplified not only at the localized source, but everywhere in space, 
and this gives the larger spreading speed.

\begin{figure}
  \includegraphics[width=\linewidth]{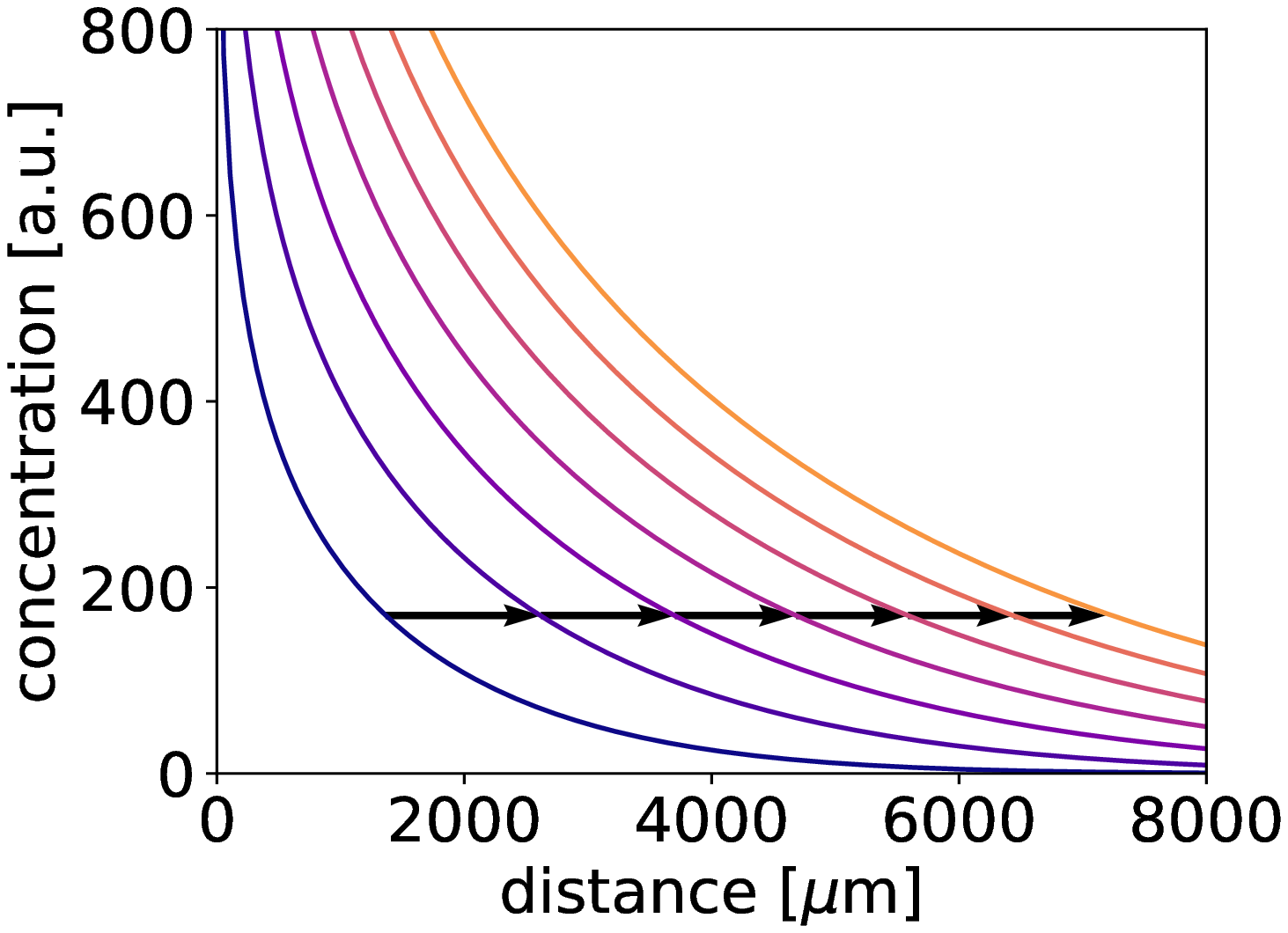}\\
  \includegraphics[width=\linewidth]{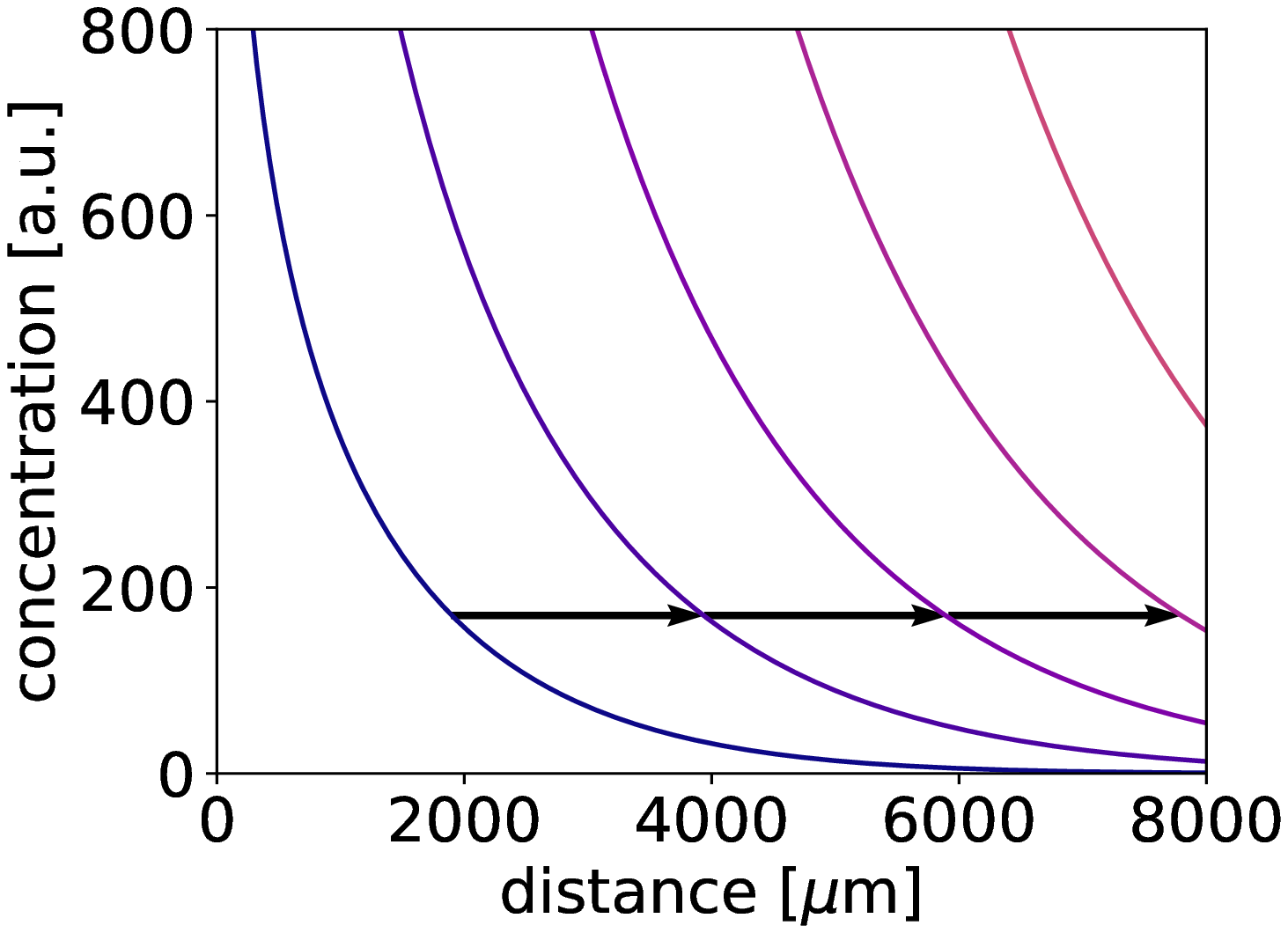}
\caption[]{
Comparison between diffusive spreading for a colony with constant production (top) and 
frontal spreading for a colony with exponentially growing production (bottom). 
The time increases inwards to outwards (dark to light) in 
steps of the doubling time of the colonies. 
The arrows indicate the distance a particular concentration level covers during
one doubling time: in the case of the steady source, the length of the arrows decreases
with time, whereas it is constant for the exponentially increasing source. Furthermore, the gradients decrease with time for the steady source but vary little for the 
exponentially growing source. The absolute
concentrations for the exponentially growing source quickly outrun those for the
constant source.
The diffusion constant in these figures is $490\, \mu$m$^2/$s and the time between
two profiles is $2.4$ h.
}
\label{front_s}
\end{figure}

To illustrate the effect of exponential growth on the spreading of the signal over time, we compare the propagation of a particular concentration level originating from an exponentially growing source to the spreading of the same concentration level produced by a constant source in Figure \ref{front_s}.
The profiles are obtained by numerical integration of (\ref{diff_2d_soln}). 
Since we do not have absolute values for the concentrations of the signaling molecules, we have to 
work with arbitrary units in the concentration. As diffusion constant we use $D=490\, \mu$m$^2/$s as estimated by \cite{Stewart:2003xy} and also used by \cite{Alberghini:2009gi}, as doubling time we work with $T_2=2.4$ h, a representative value for our model system. 
The distances from the source are given in  $\mu$m and can be compared directly with the experimental setup below.
The arrows indicate the radial displacement of the particular concentration level over a 
time interval that corresponds to the doubling time for the exponentially growing source: They are of constant lengths
for the exponential source and decrease with time for the constant source.

The preceding discussion focuses on diffusion in unbounded space.
The full solution to the problem has to take into account the finite size of the domain
with appropriate conditions near the boundaries. We assume that molecules are reflected
at the walls, so that there are no losses across the boundaries. For such 
Neumann boundary conditions the normal derivative of the concentration vanishes. 
Moreover,  we do not allow for degradation of the molecules or an absorption in the agarose, 
so that all signaling molecules are conserved, and the overall concentration will increase 
as long as the source is  active.

\begin{figure}
  \includegraphics[width=\linewidth]{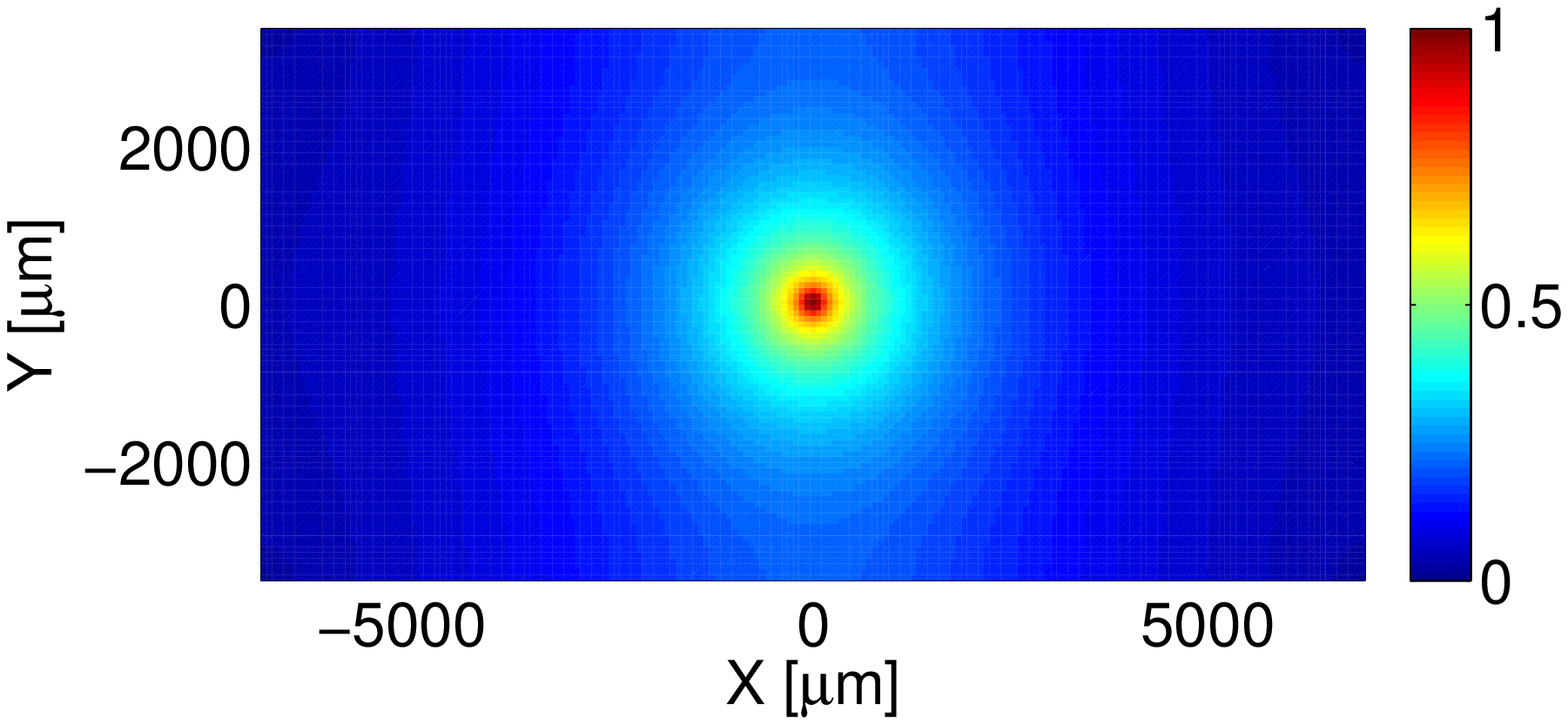}\\
  \includegraphics[width=\linewidth]{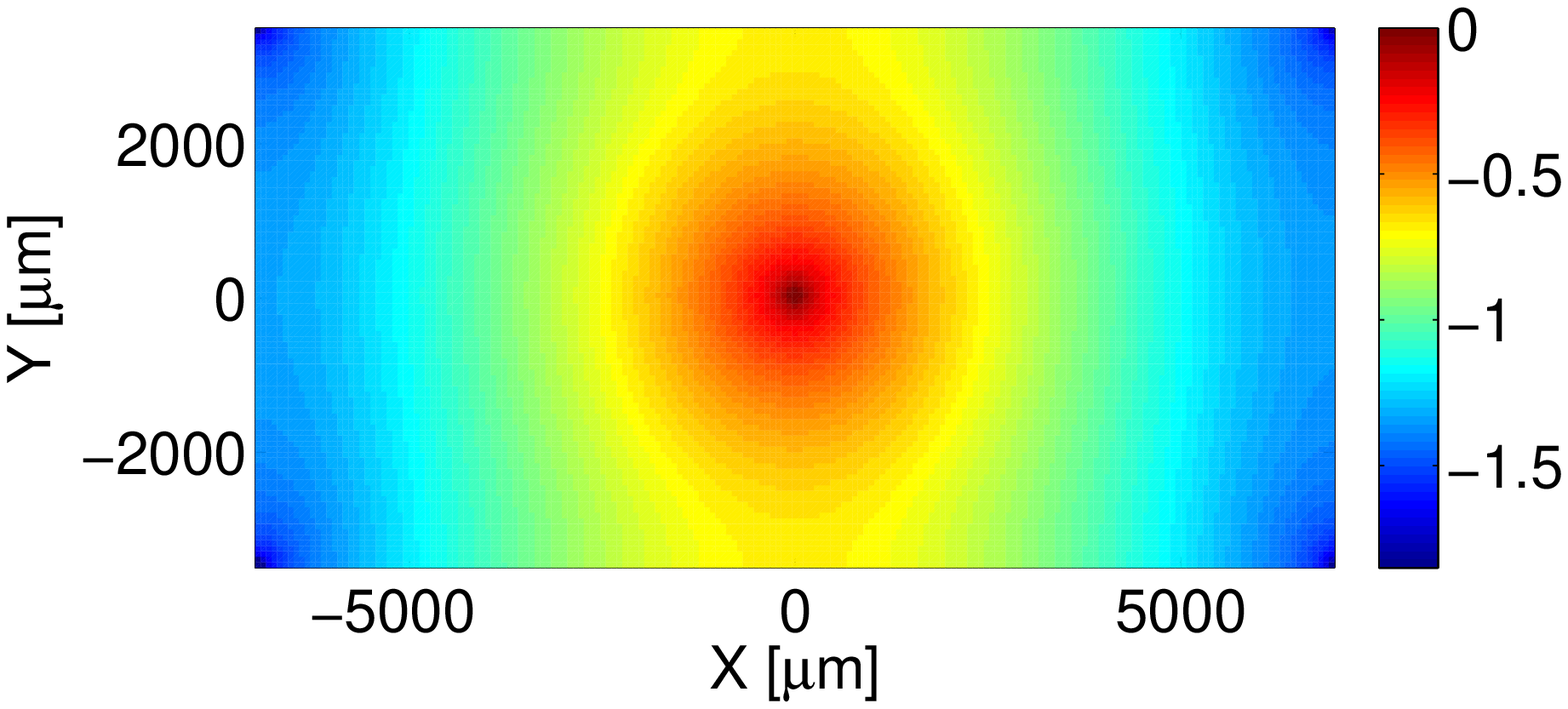}\\
\caption[]{Spatial concentration profile from a single source in the center of a rectangular pad
with reflecting boundary conditions. The concentration profiles are shown in linear (top)
and logarithmic scaling (bottom). The equidistant
spacing between steps in color reflects the exponential decay. Deviations from the 
circular symmetry are due to the rectangular domain and boundary conditions. 
}
\label{single}
\end{figure}

For the case of several sources that grow at the same rate, one can determine
the spatial variation of the profile by splitting off the exponential growth, viz.
\beq
c(\mathbf{x},t)=\tilde c (\mathbf{x}) \exp(\lambda t)\,.
\eeq
Then $\tilde c$ satisfies the time-independent diffusion equation
\beq
\lambda \tilde c = D \Delta \tilde c + \tilde q_0(\mathbf{x})\,,
\eeq
which can be solved numerically, for instance by discretizing on a square lattice. As an example, we show in Figure \ref{single} the concentration for an isolated source in the center of a rectangular domain and with reflecting boundary conditions at the walls. The deviations from a radially symmetric concentration profile are
due to the influence from the boundaries, which is larger
in the vertical direction than in the horizontal one because of the choice of a rectangular
domain:  it is $12$ mm long and $6$ mm wide, somewhat smaller than the agarose pads used in the
experiment. The other parameters are a diffusion constant of $490\, \mu m^2/s$ and 
a growth rate $\lambda=\ln 2/144$ min$^{-1}$.
Concentration profiles for several sources can be obtained by superimposing
the profiles for individual sources. An example will be given below.

The response
of the receiver colonies to this signal, which has been described by \cite{Charoenpanich:2013gp,Krol:2014jp} for \textit{S. meliloti}, can be used to detect the spreading of the front across the pad, as will be discussed in section \ref{results}. 

\begin{figure}
  \includegraphics[width=0.8\linewidth]{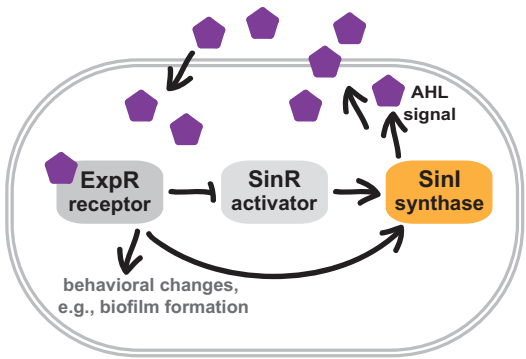}
\caption[]{Simplified illustration of the \textit{S.  meliloti} quorum sensing network. The AHL 
signaling molecules are produced by the AHL synthase SinI, spread in the environment, and are sensed by the AHL receptor ExpR. The ExpR-AHL complex (i) induces behavioral changes 
like biofilm formation, and (ii) enhances expression of the synthase gene \textit{sinI} 
in a positive feedback loop. At very high AHL concentrations, expression of \textit{sinR}, the gene 
encoding the transcriptional activator of \textit{sinI}, is repressed in a negative feedback loop. As SinR is essential for 
\textit{sinI} expression, this results in a down-regulation of the whole quorum sensing system.
 }
\label{QS_network}
\end{figure}

\section{The soil bacterium \textit{Sinorhizobium meliloti} as a model system}
\label{sino}

\textit{S. meliloti} is a Gram-negative $\alpha$-proteobacterium that engages in nitrogen fixation when 
living in symbiosis with leguminous plants 
\cite{Jones:2007dw}. 
However, it is not an obligate symbiont, but can also be found free-living in the rhizosphere 
\cite{Rinaudi:2010ko}. 
Quorum sensing plays an important role for both of these lifestyles as it contributes to 
the establishment of symbiosis
\cite{Gurich:2009hm},
and strongly stimulates extracellular matrix production, a key feature of biofilm formation 
\cite{McIntosh:2008hc}.

The \textit{S. meliloti} quorum sensing system (Figure \ref{QS_network}) 
is based on long-chained (C14-C18) acyl-homoserine lactones (AHLs) as signaling molecules \cite{Marketon:2002ev}. 
These are  produced by the AHL synthase SinI and sensed by the AHL receptor ExpR, a transcriptional regulator. 
Upon AHL binding, ExpR stimulates expression of a large number of quorum sensing target 
genes, some of which are associated with the above-mentioned changes in lifestyle. Furthermore, in a positive feedback loop, the ExpR-AHL complex stimulates expression of \textit{sinI}, the gene encoding the AHL synthase SinI, and thus upregulates signal production. However, \textit{sinI} expression can only take place in the presence of a 
second transcriptional activator, SinR. In the absence of AHLs or at low AHL concentrations, 
SinR is responsible for basal rate \textit{sinI} expression and, consequently, basal rate signal 
production. At very high AHL concentrations, expression of the \textit{sinR} gene is repressed by 
the ExpR-AHL complex, a negative feedback loop ultimately leading to a down-regulation 
of the whole quorum sensing system 
\cite{Gao:2005kp,McIntosh:2009db,Charoenpanich:2013gp}. 

\begin{figure}
 \includegraphics[width=\linewidth]{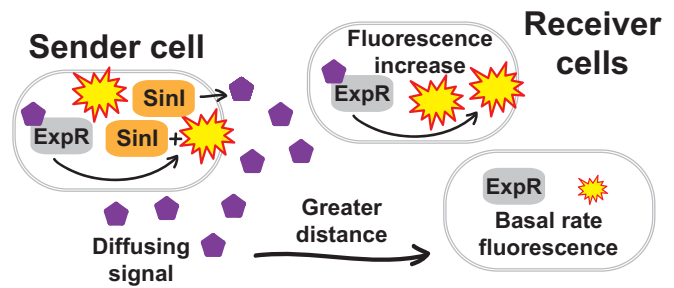}
\caption[]{Sketch of the experimental setup based on \textit{S. meliloti} strains carrying a fluorescent reporter gene fused to the \textit{sinI} promoter, thus serving as a proxy for the activity of the quorum sensing system. Sender cells (left) produce and release signaling molecules (purple pentagons) and display increased fluorescence due to the positive feedback characteristic of the quorum-sensing state. The signaling molecules diffuse across the agarose pad with receiver cells located at varying distances to the senders. When the receiver cells detect the signaling molecules, their fluorescence increases due to the same positive feedback loop (middle). Since the signal concentration strongly decreases with the distance from the senders, receiver cells at a greater distance will not experience this positive feedback and only show basal rate fluorescence (right).}
\label{setup}
\end{figure}

The experimental setup we used to explore the dynamics of the quorum sensing process between spatially separated bacterial colonies is sketched in Figure \ref{setup} (for details, see materials and methods in section \ref{methods}). Essentially, we created a setting with only one-way signal transmission based on two genetically different \textit{S. meliloti} strains: A sender strain that can both produce and sense the signal, and can thus enter the above described state of positive-feedback-related increased signal production; and a receiver strain that cannot synthesize the signaling molecules due to a partial deletion of the \textit{sinI} promotor and \textit{sinI} gene, but can still react to them. Colonies of the sender strain serve as localized sources of signaling molecules on an agarose pad prepared with defined medium, generating a spatial gradient over this essentially two-dimensional experimental field. The timing of the response observed in receiver colonies located at varying distances to the sender colonies then enables us to extract information about the dynamics with which the signaling molecules spread.

In order to make this response traceable, both strains carry a gene encoding a fluorescent protein fused to the promoter of the AHL synthase gene \textit{sinI}, i.e., the genomic region that ultimately regulates production of the signaling molecules. Therefore, both strains show basal rate fluorescence - the sender strain as it engages in basal rate signal production, and the receiver strain as it tries to generate this basal signal level, albeit it is incapable of doing so. Furthermore, both strains can show an increase in fluorescence: the sender
when engaging in increased signal production after entering the quorum-sensing state, and the receiver 
when responding to an incoming signal. 

\begin{figure}
 \includegraphics[width=0.8\linewidth]{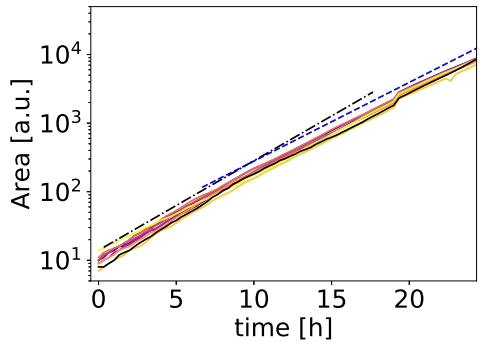}\\
\caption[]{Colony areas over time for one sender (black line) and eight receiver colonies; the respective experiment will be described in detail in section \ref{results}. Initially, all areas increase exponentially with a doubling time of about $2.4\, $h, indicated by the black dash-dotted line. After about $10\, $h, cells push on top of each other and colonies grow in height as well, so that the slope in area decreases slightly, as indicated by the blue dashed line. 
The kink in the growth curves near $20\, $h is due to a change in the microscope setting: 
As the colonies were about to grow larger than the field of view, their area and total fluorescence 
had to be determined by stitching $2\times2$ images per time frame.
}
\label{growth}
\end{figure}

Of these two strains, cell suspensions with very low optical density were spotted on the agarose pad: a single 
spot of the sender cell suspension, followed by five equidistant spots of the receiver cell suspension, 
yielding between about one dozen and eight dozen single cells on the whole pad, depending on the particular experiment. Through iterative growth and cell division, these single cells subsequently developed into large three-dimensional colonies. Colony growth and activity of the quorum sensing system in these colonies were followed via time-lapse fluorescence microscopy for approximately 24 h. During this period, colony areas grew exponentially with a mean doubling rate of $2.4\, $h for about 10 h (Figure \ref{growth}), which roughly corresponds to \textit{S. meliloti} generation times reported by \cite{Schluter:2015ax}. Subsequently, colonies became three-dimensional, and the increase in area was slightly reduced. However, no difference was observed in the growth behavior of sender and receiver strains.

\begin{figure}
  \includegraphics[width=0.8\linewidth]{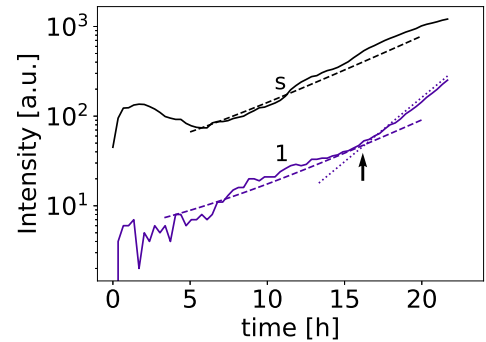}\\
\caption[]{Fluorescence intensities for a receiver colony (numbered $1$) and a sender colony labelled `s`. The dashed lines and the dotted line are obtained from fits of the form
$I(t) = a + b \exp(\lambda t)$ and are displayed after subtraction of a constant background intensity.
During the initial phase, the signal intensity of the receiver colony shows the same increase as that of the sender colony (dashed lines). For later times, the signal from colony 1 shows a transition to a steeper slope (dotted line),
with the transition point identified by the crossing of the two fits, 
as indicated by the arrow. The increase is caused by the response of the receiver colony to the AHL signal 
and marks the time when the respective threshold concentration reaches the position of the receiver colony.
}
\label{comparison}
\end{figure}

Activity from the AHL synthase promoter-fluorophore gene fusions was determined as 
mean fluorescence values, i.e., the total fluorescence 
intensity was collected over the whole colony area and then divided by the area. Based on this 
read-out, we observed different phenomena (Figure \ref{comparison}). At early time points, there is considerable variation in fluorescence levels both in sender and in receiver colonies. This variation is probably due to inaccuracies in image segmentation, and/or to fluctuations in the fraction of cells in each colony activating their AHL synthase gene promoter, as this promoter is activated heterogeneously \cite{Schluter:2015ax}. Both causes would weigh heavier the smaller the colonies are, and the lower the cell numbers. However, as this variation is limited to earlier time points, we did not explore it further.

Once inter-colony variation in fluorescence becomes negligible, fluorescence from the sender colonies is always 
significantly higher than that from receiver colonies. Thus we conclude that sender colonies enter the 
quorum-sensing state already during the first few hours of colony development, and then constantly 
produce signaling molecules at this elevated rate for most, if not all, of our observation time.

Next, we see a low but exponential increase in fluorescence from the AHL synthase gene expression reporter. 
This first increase is identical both in sender and in receiver colonies (even though absolute 
values are higher for sender than for receiver colonies, see preceding paragraph). This suggests that this first increase is unspecific, possibly originating from accumulation of the fluorophores, scattered fluorescence and fluorescence from cells within the three-dimensional colonies that are not in the focus plane of the microscope, but whose fluorescence signal is nevertheless detected by our camera. As this first increase reproducibly occurs in all colonies, we take it as the baseline for our observations.

The feature that is correlated with the spatiotemporal spreading of signaling molecules is limited to 
receiver cells only: they can show a second, more pronounced increase in fluorescence that rises well 
above the baseline. Whether or not this increase occurs at all depends on the presence of the sender 
strain - we did not observe it in a control experiment with receiver colonies growing on an agarose 
pad without any sender colonies. And if it occurs in experiments with 
both senders and receivers, the timing correlates with the distance of the respective receiver 
colony from the sender colony or colonies. Therefore, we interpret this second increase as the specific response 
of receiver colonies to incoming quorum sensing signaling molecules, namely the activation of the 
above-described positive feedback loop on AHL synthase gene expression by receptor-bound AHLs, and, 
thus, the inter-colony communication we set out to study.

\section{Results and discussion}
\label{results}

To be able to draw not only qualitative conclusions - that the timing of the receivers' reactions is indeed distance-dependent -, but to further characterise this dependence, we analyzed the data as described in Figure \ref{comparison}: Two fits were made to the fluorescence intensities of those receiver colonies that showed an AHL-dependent increase in fluorescence, one for earlier times, the second for the part with the steeper increase. The crossing of the two fits marks the onset of the receivers' quorum sensing response, and thus the arrival of the signal threshold concentration at the position of the  receiver colony. These transition times containing the temporal information were then plotted against the spatial value, either the distance of the respective colony from the sender colony, or its position on the $x$-axis (see below).

\begin{figure}
  \includegraphics[width=0.8\linewidth]{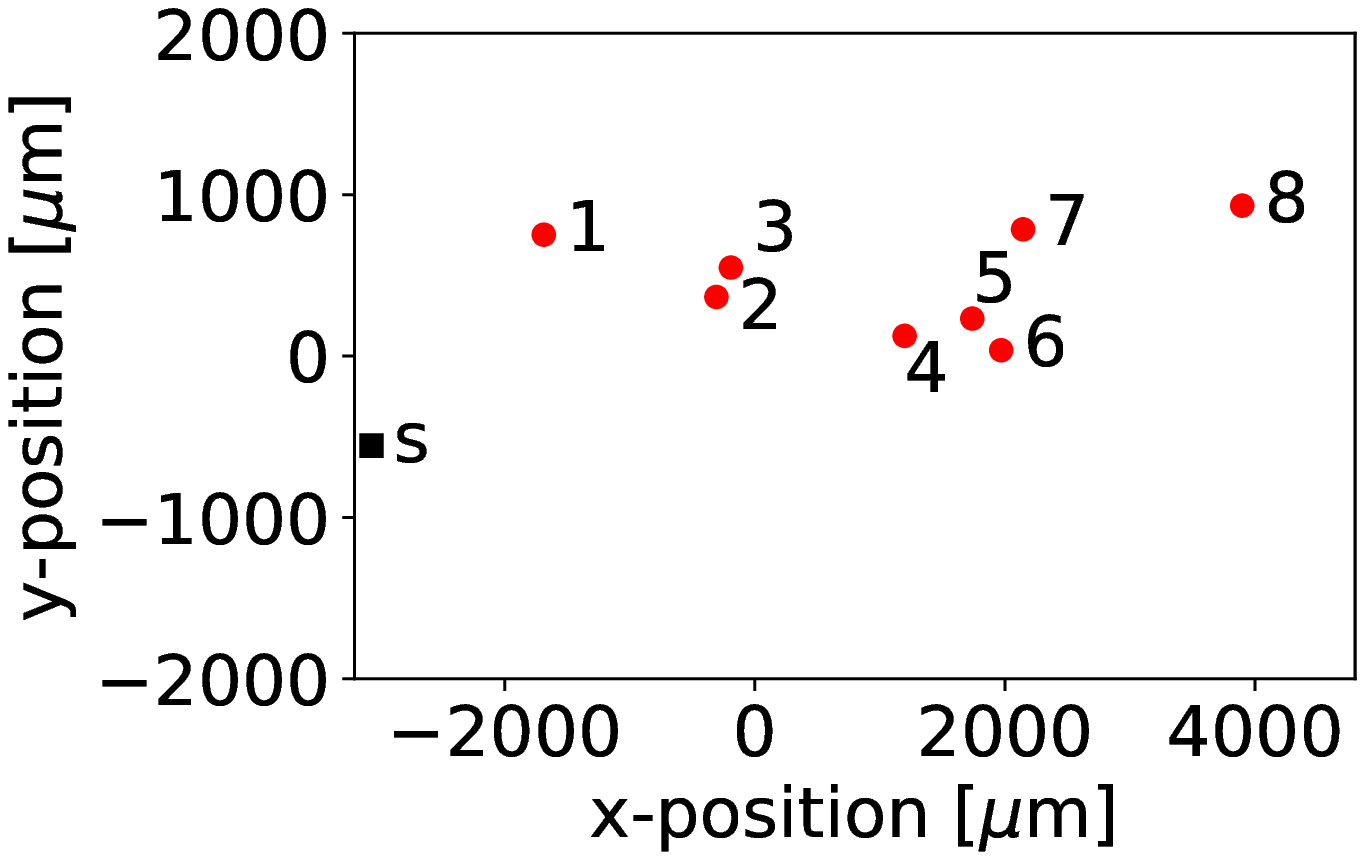}\hfill.\\
  \includegraphics[width=0.8\linewidth]{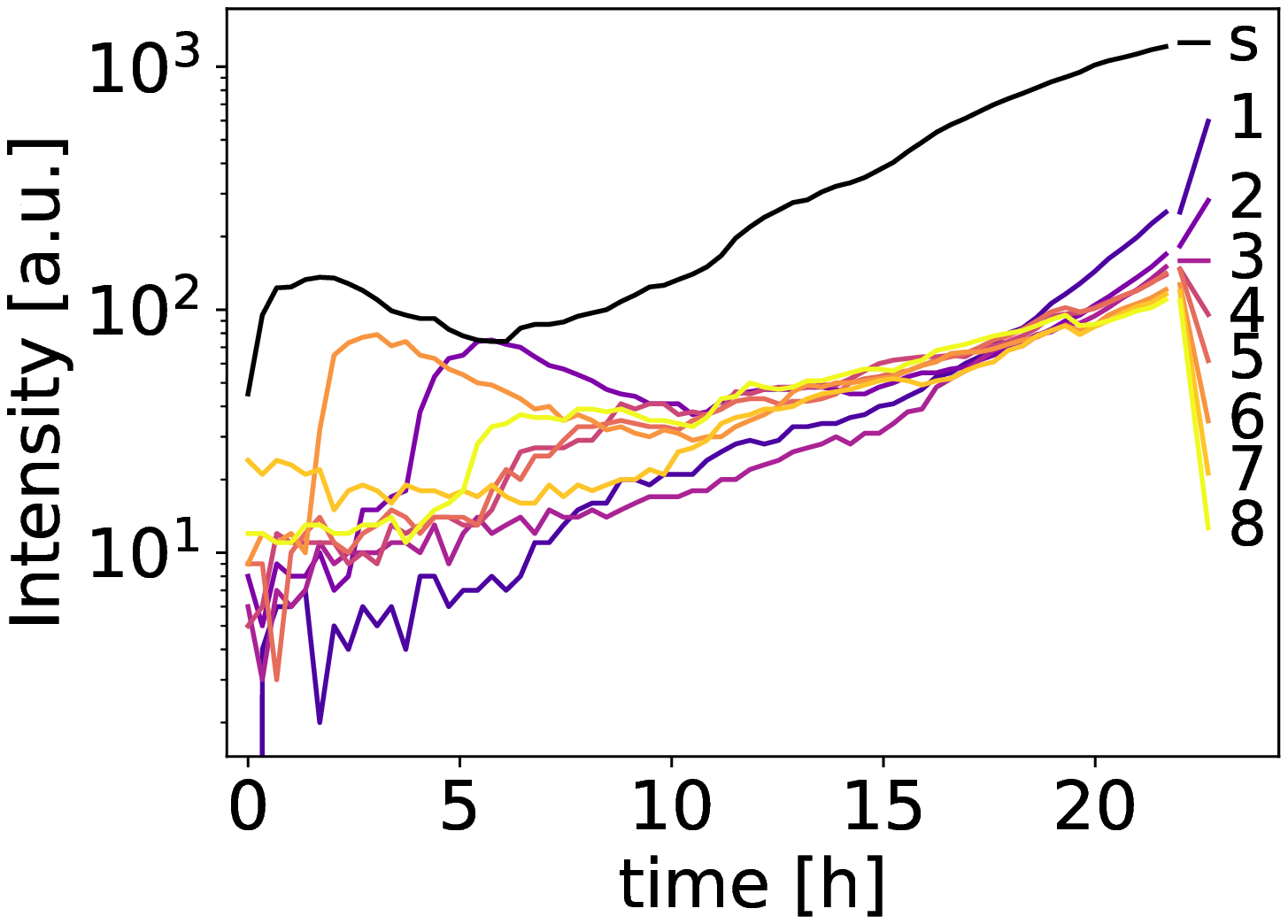}\\
  \includegraphics[width=0.8\linewidth]{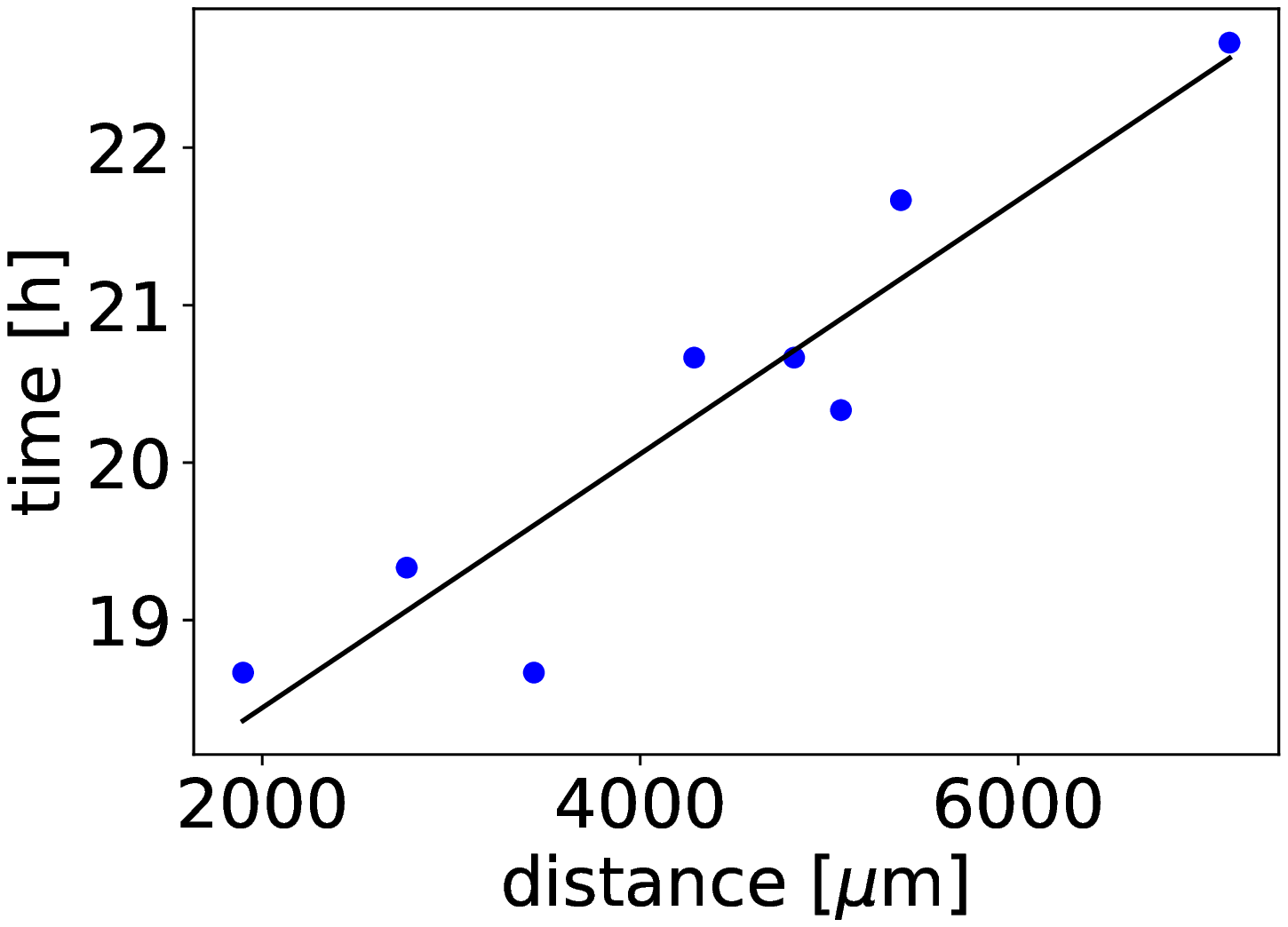}\\
\caption[]{The experiment with the single sender colony and eight receiver colonies. The top frame shows the locations of the sender colony (black square, label `s`) and the receiver colonies, numbered 1 to 8. In the middle frame, the fluorescence signals from all colonies are shown. In the bottom frame, the times at which  each fluorescence signal shows the AHL-induced increase is plotted vs. the distance of the respective receiver colony from the sender colony. The black line represents the linear regression.}
\label{plate_feb20}
\end{figure}

This is illustrated for the experiment with the single sender colony and eight receiver colonies in Figure \ref{plate_feb20}: In the top frame, the positions of the respective colonies on the agarose pad are shown. The middle frame gives the fluorescence intensities for all colonies, which already indicates the distance-dependence of the receivers' reactions, albeit only qualitatively (the curves of the receiver colonies are color-coded according to the distance from the source). The plot of the transition times against the distance from the sender colony in the bottom frame then 
clearly demonstrates that this dependence is of a linear form, as follows from the spreading of the quorum sensing signal in propagating fronts proposed by our model in section \ref{model}.  

\begin{figure}
  \includegraphics[width=0.8\linewidth]{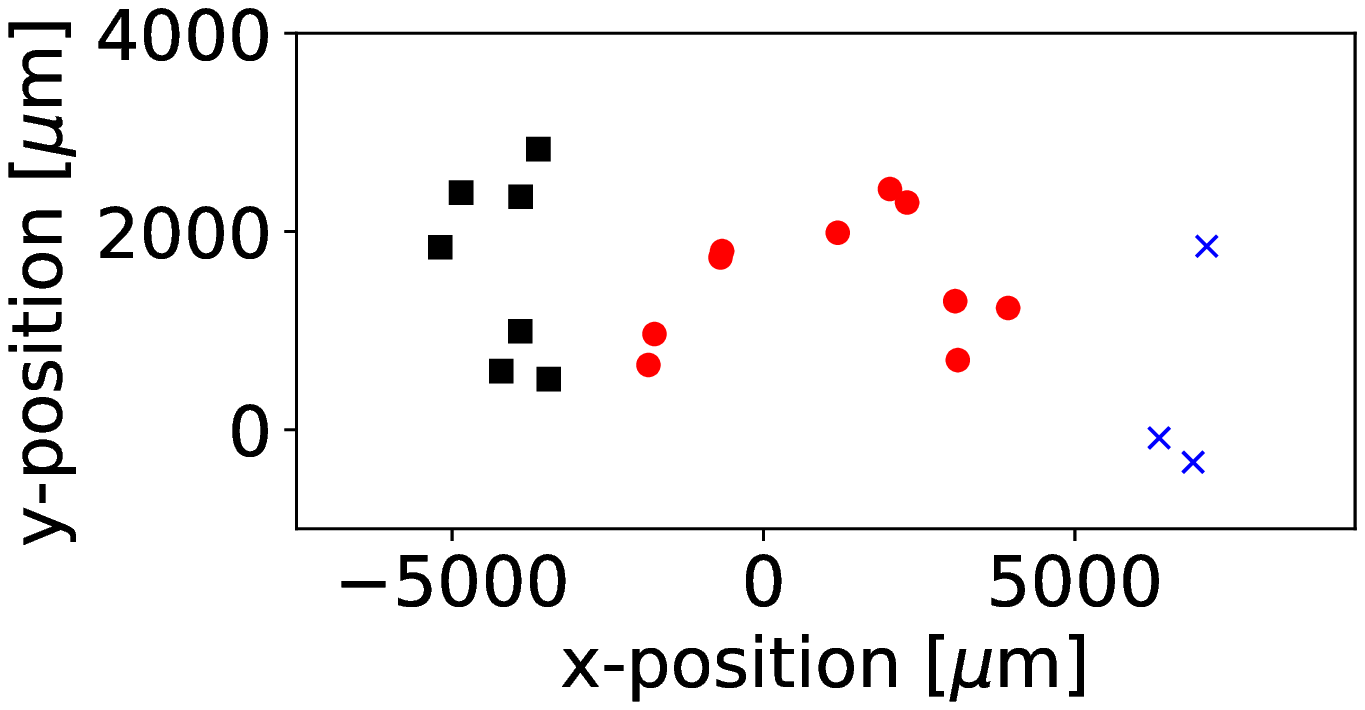}
  \includegraphics[width=0.8\linewidth]{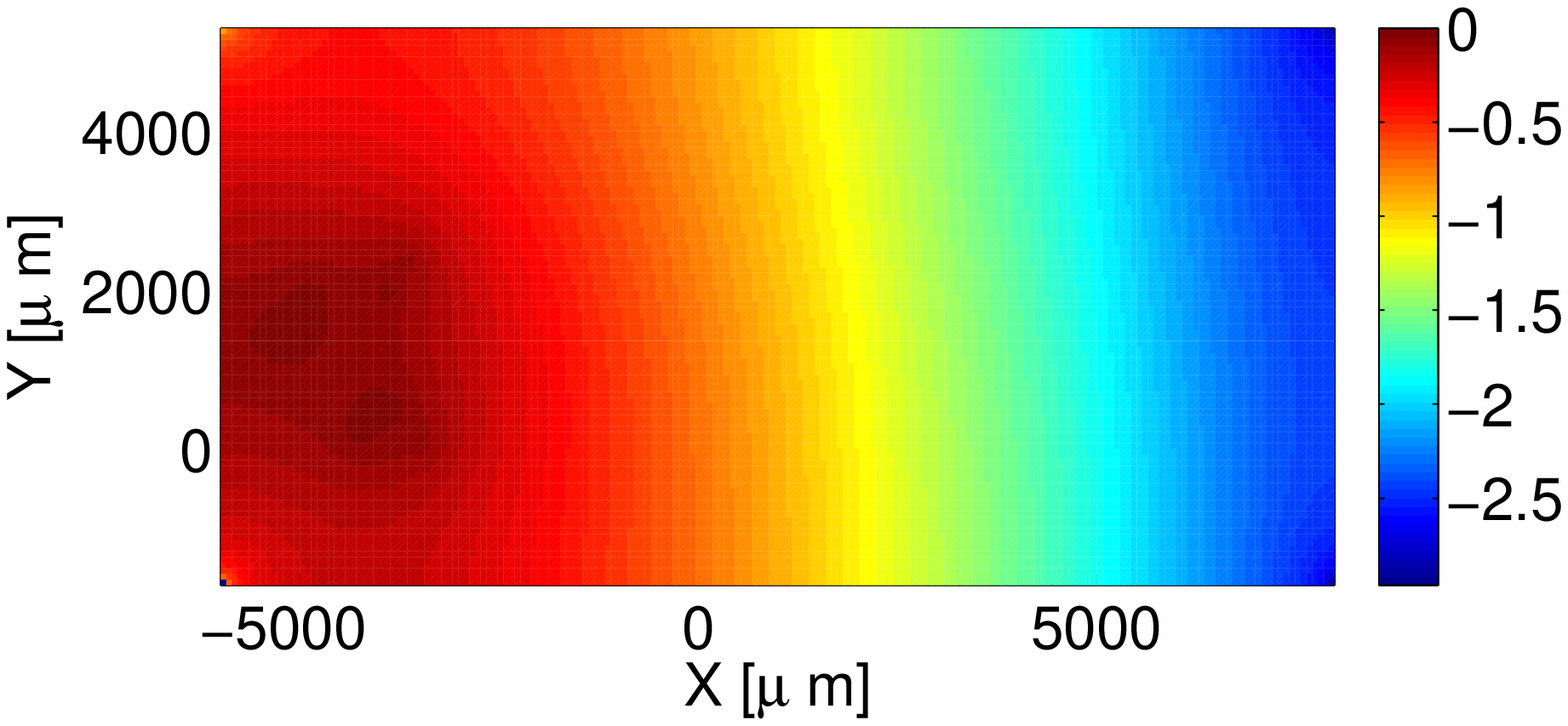}
  \includegraphics[width=0.8\linewidth]{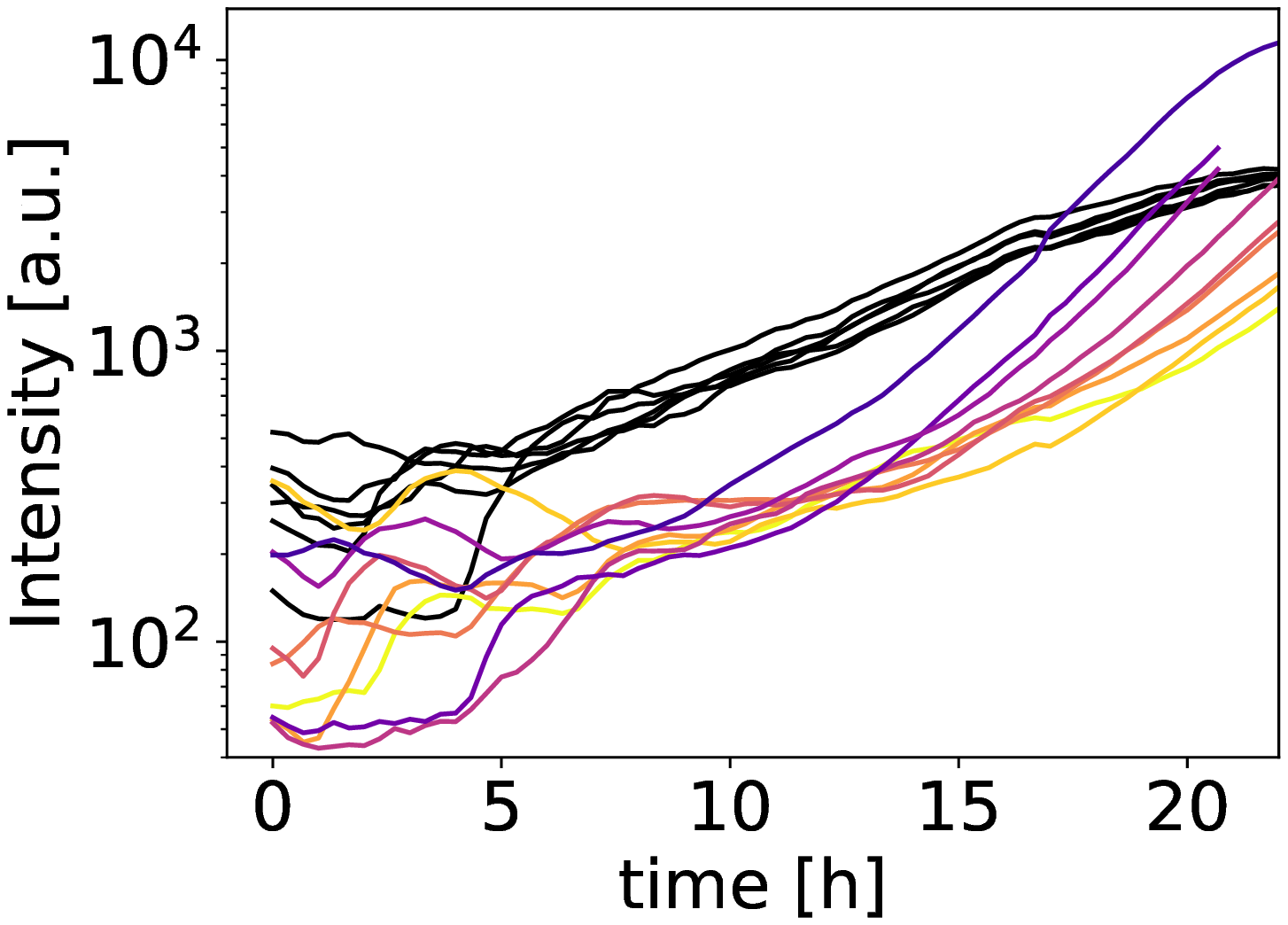}
  \includegraphics[width=0.8\linewidth]{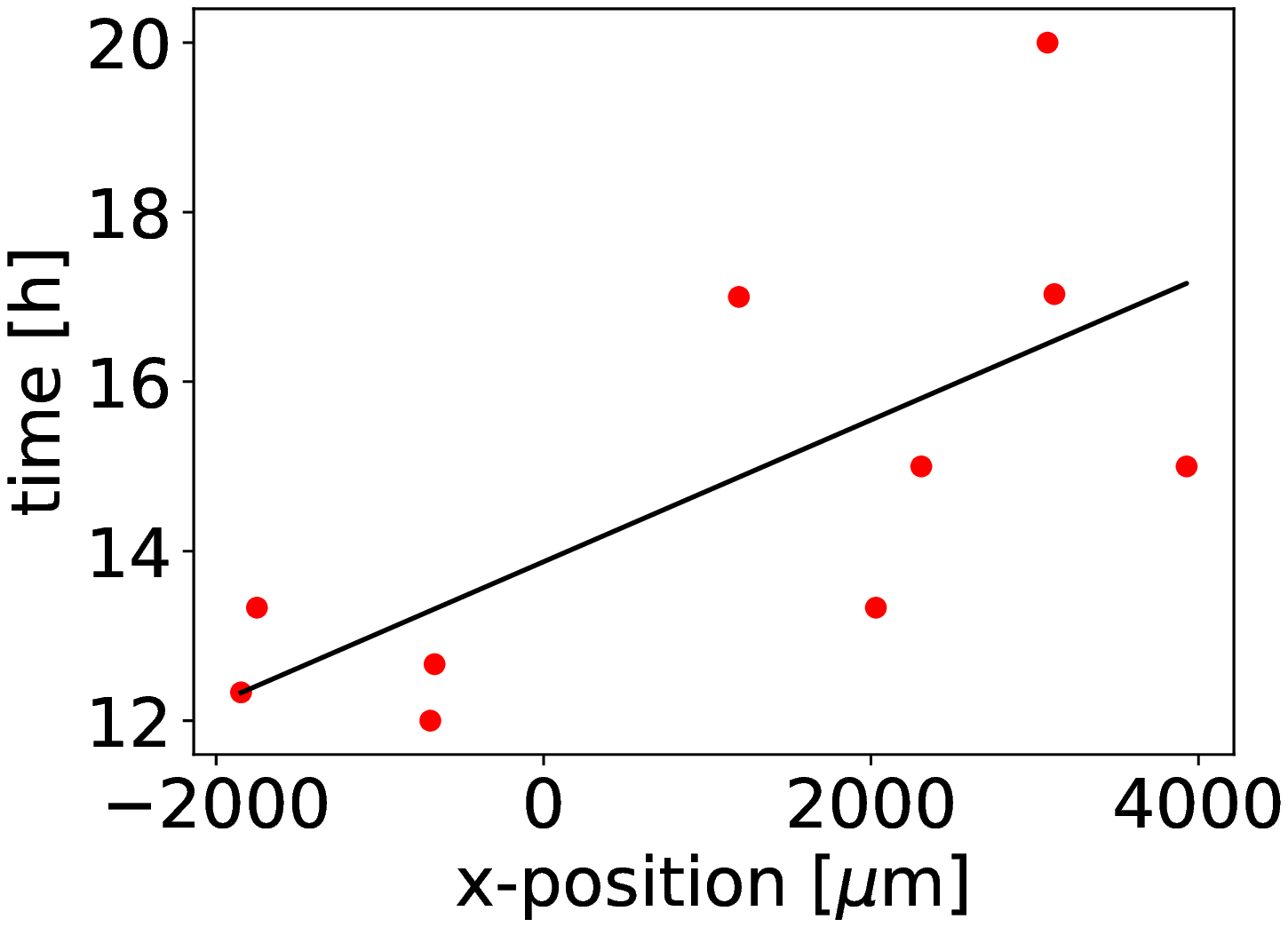}
\caption[]{An experiment with seven sender colonies. The top frame again shows the location of the sender colonies (black squares), 10 receiver colonies that do show a response to the AHL signal (red dots) and 3 further receiver colonies that do not (blue crosses). Second frame: The solution of the diffusion equation for the spatial profile in logarithmic scale
shows that to a reasonable approximation the concentration profile falls off exponentially
along the x-axis and is constant along the y-axis. Third frame: The fluorescence signals collected from all sender (black) and receiver cells (color-coded from dark to light depicting increasing distance from the source). The bottom frame shows the relation between times of transition and position of the receiver colonies, including a linear regression. 
}
\label{plate}
\end{figure}

A further experiment with seven sources is analyzed in Figure \ref{plate}. Each of these sender colonies will contribute a radial signal profile like the one shown for a single source in Figure \ref{single}. The 
combined signal concentration can be computed numerically, and the logarithmic presentation in the second frame
demonstrates that the superposition of the 
contributions from the seven sources gives rise to a profile that hardly varies along the $y$-axis of the agarose pad, but falls off exponentially along the $x$-axis, as indicated by the equidistant spacing between the color regions. Effectively, over the distances analyzed here, the seven sources 
represent a line source in the vertical direction. It is thus possible to switch to a 1-d representation of the diffusion process and, for the spatial information, to replace the distance of the receiver colonies from the sender colonies by their horizontal position. Extracting the transition points in time for the different colonies from the fluorescence signals in the third frame and correlating them with the respective $x$-positions then gives the plot in the bottom frame: The data show a larger scatter than the ones from the experiment with the single source, but the same linear spatiotemporal relationship. 

Note that the response of the receivers in the present case with seven sources occurs at earlier times compared to that of the case of the single source. We interpret this as a consequence of the difference in sender cell numbers in the two experiments: To produce the same AHL threshold level at a particular distance - e.g., the position of the nearest receiver colony -, the single sender colony needs to double about 2.8 times more often, for which it needs about 6.7 h, and this is roughly the shift between the response times in Figures \ref{plate_feb20} and \ref{plate}.

Thus, in each experiment, the absolute transition times of the receiver colonies depend on the number of sender colonies producing the signal. The absolute distances on the other hand depend on the location of the source. 
In order to combine data from different experiments, this dependence can be eliminated by shifting positions
and times such that the mean values vanish, i.e., with $t_i$ the response time at distance
$r_i$, we compute the averages $\overline t$ and $\overline r$ and determine the relative times and positions $\tilde t_i = t_i - \overline{t}$ and $\tilde r_i = r_i - \overline{r}$,
respectively. These numbers are independent of the actual number of sender cells, and of the actual location of these cells, and only give information about the dynamics of the signal propagation over the receiver-covered distance. 

\begin{figure}
  \includegraphics[width=\linewidth]{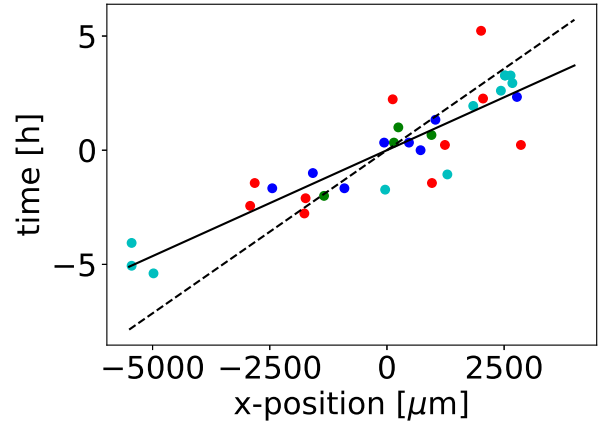}\\
\caption[]{Transition times vs. position for four data sets, the two shown in Figures \ref{plate_feb20} (blue)
and \ref{plate} (red), as well as one case with three sender colonies (dark green) and one with 68 sender colonies (cyan). 
In order to be able to compare the data in one figure, all sets have been balanced in 
position and time so that the averages are zero. The slope of a linear regression 
then still gives the velocities: the continuous line is the result from a linear regression of
all data points, with $c=1080\, \mu$m$/$h, the dashed line a relation with a 
front speed of $v=700\, \mu$m$/$h that
follows from a diffusion constant of $D= 490\,  \mu$m$^2$/s as estimated by \cite{Stewart:2003xy}.}
\label{velocity_fit}
\label{all_data}
\end{figure}

By applying such normalizations, we merged the data for the two experiments shown in Figures \ref{plate_feb20} and \ref{plate}, as well as additional data from two further experiments with three and 68 sender colonies, respectively. The fact that some of the data in the resulting Figure \ref{velocity_fit} is rather scattered might be explained as follows: For the temporal axis, the determination of exact transition points is sometimes hindered by, e.g., fluctuations in the fluorescence signal, as these can impede the calculation of the fits, or by the time-lapse character of our data acquisition where imaging might have taken place right after or right before the onset of a given transition. For the spatial axis, at least in the experiments with more than one source, the $x$-positions of the receiver colonies are only an approximation and very likely not as precise as the absolute distances given for the case with the single source - the least-scattered of our data sets. Nevertheless, the data of the four independent experiments are all distributed in a similar fashion, are all compatible with a linear increase in time with increasing distance from the source, and thus all support spreading of the quorum sensing signal in propagating fronts.

For the velocity of the front, we can determine the slopes  in linear least square fits.
Both the experiments with the single sender and the seven senders - i.e.,  Figures \ref{plate_feb20} and \ref{plate}, respectively - give a front velocity of $v=1200$ $\mu$m$/$h. 
A linear regression to the collected data in Figure \ref{velocity_fit} gives a slightly 
smaller front propagation speed of about  $v=1080\, \mu$m$/$h.
In order to connect to other data in the literature,
we use the relation between velocity, diffusion constant, and growth rate given by equation (\ref{velocity}), so that
$D = v^2/\lambda$.  With a doubling time of $2.4\,$h the growth rate $\lambda$ becomes $\lambda= 0.29\,$h$^{-1}$, which together with the front speed of $v=1080\, \mu$m$/$h gives a diffusion constant 
of about $1120\, \mu$m$^2$/s.

This value is much larger than that quoted by \cite{Basu:2005cq} in their simulations ($D= 17\, \mu$m$^2$/s),
and also larger than the value given in \cite{Ortiz:2012gq} ($D= 71\, \mu$m$^2$/s). However, Stewart \cite{Stewart:2003xy} gives estimates for diffusion constants of 490 and $720\, \mu$m$^2$/s for AHLs with chains of
12 and 4 carbon atoms, respectively,  \cite{Alberghini:2009gi} use the value of $490\, \mu$m$^2$/s for their calculations, and \cite{Dilanji:2012ds} use $720\,  \mu$m$^2$/s.
Furthermore, as discussed by \cite{Dilanji:2012ds}, diffusion of AHLs is influenced by, e.g., the length of the acyl side chains, and the stability of the molecules is strongly affected by the pH. 

The relation we use to deduce the diffusion constant contains the square of the front velocity, so that uncertainties
in the front velocity are amplified considerably. On the other hand, assuming the
same growth rate, but a diffusion constant that is 16 times smaller (i.e., going down from
$1600\, \mu$m$^2$/s to $100\, \mu$m$^2$/s) would reduce the front speed 
by a factor of 4, down to 320 $\mu$m$/$h. Accordingly, the time to cover the distance over 
which the responding receiver colonies are spread in our experiments (about $6000\, \mu$m) would increase 
from the observed interval of about $5\,$h to $20\,$ h, close to the total run time of our 
experiment. Alternatively, in order to obtain the front velocity deduced from
Figure \ref{velocity_fit} with such a small diffusion constant would require a 
growth rate that would have to be 16 times faster than the one we used for our calculations,
which is well outside the range of uncertainty of our experiments. On the other hand, a diffusion constant of $490\,  \mu$m$^2$/s as estimated by \cite{Stewart:2003xy} gives a front speed of $700\, \mu$m$/$h, corresponding to the dotted line in Figure \ref{velocity_fit}: It has a steeper slope, but still runs through
most of the data, and it would also be compatible with the time interval of about 5 h over which the
receiver colonies respond to the signal. Thus, while the origin of the variations between the 
different experiments and estimates remains unclear, we conclude that front speeds 
in the range of $700-1100\,\mu/$h, with diffusion constants in the range of 
$500-1150\, \mu$m$^2/$s are compatible with the present observations.

\section{Concluding remarks}
\label{conclusion}
Without nonlinear effects, diffusion gives rise to concentration profiles with poorly defined gradients, so that receiver cells that react to a certain signal concentration will have a large variability in their response in space and time. However, as we have shown here, the combination of exponentially growing sender colonies that release an exponentially
increasing number of signaling molecules together with the diffusive spreading of the molecules creates
a concentration profile where levels of constant concentration spread in a front-like fashion:
The speed and the slopes of the profiles are constant as in many other examples of 
propagating fronts. The front speed $v=\sqrt{D\lambda}$ is given by a combination of diffusion constant $D$ and growth rate $\lambda$. For a fixed time, the concentration profile varies exponentially in space, with a characteristic length $\ell = \sqrt{D/\lambda}$. 

Observations on \textit{S. melioti} are used to test the theoretical predictions. Exponentially growing sender colonies release signaling molecules that spread over the agarose pad and trigger responses in receiver colonies that themselves cannot produce AHLs. The response is detected by an increase of fluorescence signal, and the time of response correlates linearly with the distance from the source.

The exponential variation of the signal concentration in space could not be verified directly; 
this would require either a direct measurement of AHL concentrations in agarose pads, or receiver colonies with different thresholds for the AHL response. But even with the single signal threshold level displayed by our receiver strain, the quorum sensing response of \textit{S. meliloti} to AHLs offers interesting possibilities: It sets in at low concentrations in the nanomolar range, but is blocked again at higher concentrations by a negative feedback loop \cite{Charoenpanich:2013gp,Krol:2014jp}. Combined with the homogeneously decaying profile proposed by our model, this regulatory network architecture should 
create an outwards-moving ring around producer colonies where receiver colonies will respond. 
This feature could be a mechanism for the creation of ringlike spatial patterns, similar to the ones discussed by \cite{Basu:2005cq}.

Another interesting consideration resulting from the work presented here is the following: The expression for the front speed $v=\sqrt{D\lambda}$ contains the growth rate as one of the factors. One can therefore expect that systems that grow more rapidly will also spread their signals more rapidly, and conversely if the growth rate is lower. For colonies producing and responding to the same signaling molecules, this dependence on the growth rate carries an additional means of sensing and reacting to local environmental conditions: Since regions with higher growth rates presumably have better nutrient supply and growth conditions, colonies growing here might 
dominate the local signal landscape, and if bacteria in the vicinity can follow the signal gradient, they might thus be attracted 
to more favorable environments. The linear spreading and the well defined gradients would make this signal much easier to follow than nutrient gradients that only spread diffusively.

\section{Materials and Methods}
\label{methods}

\subsection{Bacterial strains and growth conditions}
The strains used in this study were generated using standard genetic techniques and grown applying 
standard laboratory practice 
\cite{McIntosh:2009db,Dohlemann:2017xy}.
After cloning in \textit{Escherichia coli} DH5$\alpha$, final constructs were verified by DNA sequencing, and if subsequent homologous recombination in \textit{S. meliloti} was involved, the
resulting strains were again verified by DNA sequencing. 

For the sender strain the AHL synthase gene promoter including the downstream 
native ribosome binding site and the 
first 27 base pairs of the \textit{sinI} gene, the \textit{mVenus} gene 
\cite{Nagai:2002wb}
including stop codon, and the 
AHL synthase gene \textit{sinI} with its native ribosome binding site were inserted into the plasmid pK18mobsacB 
\cite{Schaefer:1994ax}
in \textit{E. coli} DH5$\alpha$. This construct was transferred to \textit{S. meliloti} strain Sm2B3001 \cite{Bahlawane:2008dg} by \textit{E. coli} S17-1-mediated conjugation \cite{Krol:2014jp}. Double recombinants carrying the transcriptional fusion at the chromosomal \textit{sinI} locus were selected on LB \cite{Krol:2014jp} agar containing 10\% sucrose. 

Receiver constructs were based on the same AHL synthase promoter sequence as the sender strain. 
For the experiment with seven sender colonies, this promoter sequence was fused to the \textit{mCherry} 
gene 
\cite{Shaner:2004ax}
on the low copy plasmid pPHU231 \cite{Hubner:1991uy};
the plasmid also 
carried a second copy of the same promoter fused to \textit{cerulean} \cite{Rizzo:2004hz}
and the promoter 
of \textit{S. meliloti} gene 
\textit{SMc00877} fused to \textit{mVenus}, 
both of which were not analyzed for this study. By \textit{E. coli} S17-1-mediated conjugation, the final construct was transferred to \textit{S. meliloti} Sm2B4001 \cite{McIntosh:2009db} carrying a deletion starting 102 base pairs upstream of the \textit{sinI} ATG and including the first 66 base pairs of the \textit{sinI} gene. For all other experiments including the receiver-only control, the 
AHL synthase promoter sequence followed by the \textit{mVenus} gene was inserted into the suicide 
plasmid pK18mob2\_Km \cite{Schaefer:1994ax}.
The final construct was transferred to \textit{S. meliloti} 
Sm2B4001 via \textit{E. coli} S17-1-mediated conjugation. Recombinants carried the plasmid integrated at the chromosomal \textit{sinI} promoter locus. 

For most experiments, the \textit{sinI} promoter-\textit{mVenus} strains also carried a synthetic version of the \textit{trp} 
promoter  regulating expression of the \textit{mCherry} 
gene as a constitutive reporter. For the sender, this construct 
was cloned into the suicide plasmid pK18mob2\_Km also carrying a 
\textit{wgeA} promoter-\textit{cerulean} fusion - the \textit{wgeA} promoter drives expression of genes essential for exopolysaccharide production and served as the integration site for the plasmid. The receiver strain with the \textit{sinI} promoter-\textit{mVenus} fusion carried only the synthetic \textit{trp} promoter-\textit{mCherry} 
fusion cloned into the single copy plasmid pABC1mob \cite{Dohlemann:2017xy}.

Starter cultures for time-lapse microscopy were grown in TY medium \cite{Krol:2014jp} to an optical density measured at a wavelength of 600 nm (OD$_{600}$) of around 0.8. After harvest, cells were washed three times in an equal volume of MOPS-buffered medium \cite{Zhan:1991te}. Cell density was then adjusted to an OD$_{600}$ of  0.000005, 0.00001, 0.000025 or 0.00025, depending on the particular experiment. Agarose pads were prepared in 
17x28 mm Frame Seal in situ PCR and hybridization slide chambers (Biorad) with MOPS-buffered  
medium containing 2 mM phosphate and 1.2 \% agarose. Corners were cut off to create air reservoirs.
Slightly off the middle of the agarose pads, $0.3\,\mu$l of the sender cell suspension were spotted, 
followed by five spots of the same volume of receiver cell suspensions at 2, 4, 6, 8 and 10 mm 
distance to the sender spot along a horizontal line. For the control experiment, only receiver cell 
suspension was spotted mimicking the original experimental setup.

\subsection{Time-lapse fluorescence microscopy}
Time-lapse fluorescence microscopy was performed with an Eclipse Ti-E inverse research microscope (Nikon) equipped with a Plan Apo $\lambda$ 100x/1.45 oil objective (Nikon) in an 
incubation chamber set to 30 $^{\circ}$C. Individual \textit{S. meliloti} cells on the agarose pad were searched for 
using the live imaging mode of the NIS Elements Advanced Research software 
version 4.13 (Nikon). x, y and z coordinates of the cells were recorded in the ND Acquisition module of 
the same software, and phase contrast and fluorescence images of the respective colonies were 
automatically taken every 20 minutes using an IXON X3885 camera (Andor, Oxford Instruments) over 
a period of at least 24 h. When colonies were about to leave the field of vision of the camera 
after about 17 h, the 2x2 Large Image function of the ND Acquisition module was used 
to further follow colony expansion. After the time lapse was stopped, an overview of the agarose pad was produced with the Scan Large Image function and a Plan Fluor 4x/0.13 objective (Nikon). 

Fluorophore excitation was carried out with an Intensilight Hg Precentered Fiber 
Illuminator (Nikon). Specific interference and absorption filter sets were applied for 
mCherry and mVenus fluorescent proteins \cite{Schluter:2015ax}.
2x2 binning was used to reduce excitation intensity and 
exposure time, and, thus, phototoxicity. Electron-Multiplying (EM) gain was set to 30 for mVenus and
10 for mCherry; conversion gain was always set to 1. For each channel, 
excitation intensity and exposure time were then selected to assure optimal illumination.

\subsection{Image analysis}
Image analysis was performed using the General Analysis module of the NIS Elements Advanced 
Research software version 4.5. Binary layer construction, i.e., determination of colony perimeters, 
was performed on phase contrast images, whenever possible combined with 
quorum-sensing-unrelated fluorescence images from the \textit{trp} 
promoter-\textit{mCherry} fusion. 
Based on these colony perimeters, colony area and mean fluorescence values, i.e., the ratio of 
total fluorescence intensity per area, were determined. From these mean fluorescence values, background fluorescence was subtracted.

\section{Acknowledgements}
We thank Gabriele Malengo for discussing imaging conditions, Stephan Ringshandl for initial help with image analysis, Yasmin Hengster
for the preparation of some of the parameter fits and figures, and Peter Lenz, Knut Drescher and Moritz Linkmann 
for helpful comments on the problem and the manuscript.
This work was supported by the priority program SPP 1617 (German Research Foundation) and the LOEWE Program of the State of Hesse (SYNMIKRO). 


\begin{thebibliography}{45}%
\makeatletter
\providecommand \@ifxundefined [1]{%
 \@ifx{#1\undefined}
}%
\providecommand \@ifnum [1]{%
 \ifnum #1\expandafter \@firstoftwo
 \else \expandafter \@secondoftwo
 \fi
}%
\providecommand \@ifx [1]{%
 \ifx #1\expandafter \@firstoftwo
 \else \expandafter \@secondoftwo
 \fi
}%
\providecommand \natexlab [1]{#1}%
\providecommand \enquote  [1]{``#1''}%
\providecommand \bibnamefont  [1]{#1}%
\providecommand \bibfnamefont [1]{#1}%
\providecommand \citenamefont [1]{#1}%
\providecommand \href@noop [0]{\@secondoftwo}%
\providecommand \href [0]{\begingroup \@sanitize@url \@href}%
\providecommand \@href[1]{\@@startlink{#1}\@@href}%
\providecommand \@@href[1]{\endgroup#1\@@endlink}%
\providecommand \@sanitize@url [0]{\catcode `\\12\catcode `\$12\catcode
  `\&12\catcode `\#12\catcode `\^12\catcode `\_12\catcode `\%12\relax}%
\providecommand \@@startlink[1]{}%
\providecommand \@@endlink[0]{}%
\providecommand \url  [0]{\begingroup\@sanitize@url \@url }%
\providecommand \@url [1]{\endgroup\@href {#1}{\urlprefix }}%
\providecommand \urlprefix  [0]{URL }%
\providecommand \Eprint [0]{\href }%
\providecommand \doibase [0]{http://dx.doi.org/}%
\providecommand \selectlanguage [0]{\@gobble}%
\providecommand \bibinfo  [0]{\@secondoftwo}%
\providecommand \bibfield  [0]{\@secondoftwo}%
\providecommand \translation [1]{[#1]}%
\providecommand \BibitemOpen [0]{}%
\providecommand \bibitemStop [0]{}%
\providecommand \bibitemNoStop [0]{.\EOS\space}%
\providecommand \EOS [0]{\spacefactor3000\relax}%
\providecommand \BibitemShut  [1]{\csname bibitem#1\endcsname}%
\let\auto@bib@innerbib\@empty
\bibitem [{\citenamefont {Braun}\ \emph {et~al.}(1994)\citenamefont {Braun},
  \citenamefont {Wissing}, \citenamefont {Sch{\"a}fer},\ and\ \citenamefont
  {Hirsch}}]{Braun:1994tz}%
  \BibitemOpen
  \bibfield  {author} {\bibinfo {author} {\bibfnamefont {H A}\ \bibnamefont
  {Braun}}, \bibinfo {author} {\bibfnamefont {K}\ \bibnamefont {Wissing}},
  \bibinfo {author} {\bibfnamefont {K}\ \bibnamefont {Sch{\"a}fer}}, \ and\
  \bibinfo {author} {\bibfnamefont {M~C}\ \bibnamefont {Hirsch}},\
  }\bibfield  {title} {\enquote {\bibinfo {title} {{Oscillation and noise
  determine signal transduction in shark multimodal sensory cells}},}\
  }\href@noop {} {\bibfield  {journal} {\bibinfo  {journal} {Nature}\ }\textbf
  {\bibinfo {volume} {367}},\ \bibinfo {pages} {270--273} (\bibinfo {year}
  {1994})}\BibitemShut {NoStop}%
\bibitem [{\citenamefont {Neiman}\ \emph {et~al.}(1999)\citenamefont {Neiman},
  \citenamefont {Pei}, \citenamefont {Russell}, \citenamefont {Wojtenek},
  \citenamefont {Wilkens}, \citenamefont {Moss}, \citenamefont {Braun},
  \citenamefont {Huber},\ and\ \citenamefont {Voigt}}]{Neiman:1999bh}%
  \BibitemOpen
  \bibfield  {author} {\bibinfo {author} {\bibfnamefont {A}\
  \bibnamefont {Neiman}}, \bibinfo {author} {\bibfnamefont {X}\ \bibnamefont
  {Pei}}, \bibinfo {author} {\bibfnamefont {D}\ \bibnamefont {Russell}},
  \bibinfo {author} {\bibfnamefont {W}\ \bibnamefont {Wojtenek}},
  \bibinfo {author} {\bibfnamefont {L}\ \bibnamefont {Wilkens}}, \bibinfo
  {author} {\bibfnamefont {F}\ \bibnamefont {Moss}}, \bibinfo {author}
  {\bibfnamefont {H~A}\ \bibnamefont {Braun}}, \bibinfo {author} {\bibfnamefont
  {M~T}\ \bibnamefont {Huber}}, \ and\ \bibinfo {author} {\bibfnamefont
  {K}~\bibnamefont {Voigt}},\ }\bibfield  {title} {\enquote {\bibinfo {title}
  {{Synchronization of the Noisy Electrosensitive Cells in the Paddlefish}},}\
  }\href@noop {} {\bibfield  {journal} {\bibinfo  {journal} {Phys Rev Lett}\
  }\textbf {\bibinfo {volume} {82}},\ \bibinfo {pages} {660--663} (\bibinfo
  {year} {1999})}\BibitemShut {NoStop}%
\bibitem [{\citenamefont {Braun}\ \emph {et~al.}(1997)\citenamefont {Braun},
  \citenamefont {Sch{\"a}fer}, \citenamefont {Voigt}, \citenamefont {Peters},
  \citenamefont {Bretschneider}, \citenamefont {Pei}, \citenamefont {Wilkens},\
  and\ \citenamefont {Moss}}]{Braun:1997tn}%
  \BibitemOpen
  \bibfield  {author} {\bibinfo {author} {\bibfnamefont {H~A}\ \bibnamefont
  {Braun}}, \bibinfo {author} {\bibfnamefont {K}\ \bibnamefont
  {Sch{\"a}fer}}, \bibinfo {author} {\bibfnamefont {K}\ \bibnamefont
  {Voigt}}, \bibinfo {author} {\bibfnamefont {B}\ \bibnamefont {Peters}},
  \bibinfo {author} {\bibfnamefont {F}\ \bibnamefont {Bretschneider}},
  \bibinfo {author} {\bibfnamefont {X}\ \bibnamefont {Pei}}, \bibinfo
  {author} {\bibfnamefont {L}\ \bibnamefont {Wilkens}}, \ and\ \bibinfo
  {author} {\bibfnamefont {F}\ \bibnamefont {Moss}},\ }\bibfield  {title}
  {\enquote {\bibinfo {title} {{Low-Dimensional Dynamics in Sensory Biology 1:
  Thermally Sensitive Electroreceptors of the Catfish}},}\ }\href@noop {}
  {\bibfield  {journal} {\bibinfo  {journal} {Journal of Computational
  Neuroscience}\ }\textbf {\bibinfo {volume} {4}},\ \bibinfo {pages} {335--347}
  (\bibinfo {year} {1997})}\BibitemShut {NoStop}%
\bibitem [{\citenamefont {Braun}\ \emph {et~al.}(1999)\citenamefont {Braun},
  \citenamefont {Dewald}, \citenamefont {Sch{\"a}fer}, \citenamefont {Voigt},
  \citenamefont {Pei}, \citenamefont {Dolan},\ and\ \citenamefont
  {Moss}}]{Braun:1999ta}%
  \BibitemOpen
  \bibfield  {author} {\bibinfo {author} {\bibfnamefont {H~A}\ \bibnamefont
  {Braun}}, \bibinfo {author} {\bibfnamefont {M}\ \bibnamefont {Dewald}},
  \bibinfo {author} {\bibfnamefont {M}\ \bibnamefont {Sch{\"a}fer}},
  \bibinfo {author} {\bibfnamefont {K}\ \bibnamefont {Voigt}}, \bibinfo
  {author} {\bibfnamefont {X}\ \bibnamefont {Pei}}, \bibinfo {author}
  {\bibfnamefont {Kevin}\ \bibnamefont {Dolan}}, \ and\ \bibinfo {author}
  {\bibfnamefont {F}\ \bibnamefont {Moss}},\ }\bibfield  {title} {\enquote
  {\bibinfo {title} {{Low-Dimensional Dynamics in Sensory Biology 2: Facial
  Cold Receptors of the Rat}},}\ }\href@noop {} {\bibfield  {journal} {\bibinfo
   {journal} {Journal of Computational Neuroscience}\ }\textbf {\bibinfo
  {volume} {7}},\ \bibinfo {pages} {17--32} (\bibinfo {year}
  {1999})}\BibitemShut {NoStop}%
\bibitem [{\citenamefont {Braun}\ \emph {et~al.}(2000)\citenamefont {Braun},
  \citenamefont {Eckhardt}, \citenamefont {Braun},\ and\ \citenamefont
  {Huber}}]{Braun:2000hc}%
  \BibitemOpen
  \bibfield  {author} {\bibinfo {author} {\bibfnamefont {W}\
  \bibnamefont {Braun}}, \bibinfo {author} {\bibfnamefont {B}\ \bibnamefont
  {Eckhardt}}, \bibinfo {author} {\bibfnamefont {H~A}\ \bibnamefont
  {Braun}}, \ and\ \bibinfo {author} {\bibfnamefont {M}\ \bibnamefont
  {Huber}},\ }\bibfield  {title} {\enquote {\bibinfo {title} {{Phase-space
  structure of a thermoreceptor}},}\ }\href@noop {} {\bibfield  {journal}
  {\bibinfo  {journal} {Phys Rev E}\ }\textbf {\bibinfo {volume} {62}},\
  \bibinfo {pages} {6352--6360} (\bibinfo {year} {2000})}\BibitemShut {NoStop}%
\bibitem [{\citenamefont {Feudel}\ \emph {et~al.}(2000)\citenamefont {Feudel},
  \citenamefont {Neiman}, \citenamefont {Pei}, \citenamefont {Wojtenek},
  \citenamefont {Braun}, \citenamefont {Huber},\ and\ \citenamefont
  {Moss}}]{Feudel:2000hs}%
  \BibitemOpen
  \bibfield  {author} {\bibinfo {author} {\bibfnamefont {U}\ \bibnamefont
  {Feudel}}, \bibinfo {author} {\bibfnamefont {A}\ \bibnamefont
  {Neiman}}, \bibinfo {author} {\bibfnamefont {X}\ \bibnamefont {Pei}},
  \bibinfo {author} {\bibfnamefont {W}\ \bibnamefont {Wojtenek}},
  \bibinfo {author} {\bibfnamefont {H}\ \bibnamefont {Braun}}, \bibinfo
  {author} {\bibfnamefont {M}\ \bibnamefont {Huber}}, \ and\ \bibinfo
  {author} {\bibfnamefont {F}\ \bibnamefont {Moss}},\ }\bibfield  {title}
  {\enquote {\bibinfo {title} {{Homoclinic bifurcation in a Hodgkin--Huxley
  model of thermally sensitive neurons}},}\ }\href@noop {} {\bibfield
  {journal} {\bibinfo  {journal} {Chaos}\ }\textbf {\bibinfo {volume} {10}},\
  \bibinfo {pages} {231--239} (\bibinfo {year} {2000})}\BibitemShut {NoStop}%
\bibitem [{\citenamefont {Keener}\ and\ \citenamefont
  {Sneyd}(2008)}]{Keener:2008xy}%
  \BibitemOpen
  \bibfield  {author} {\bibinfo {author} {\bibfnamefont {J}~\bibnamefont
  {Keener}}\ and\ \bibinfo {author} {\bibfnamefont {J}~\bibnamefont {Sneyd}},\
  }\href@noop {} {\emph {\bibinfo {title} {Mathematical Physiology I: Cellular
  Physiology}}}\ (\bibinfo  {publisher} {Springer},\ \bibinfo {year}
  {2008})\BibitemShut {NoStop}%
\bibitem [{\citenamefont {Prindle}\ \emph {et~al.}(2015)\citenamefont
  {Prindle}, \citenamefont {Liu}, \citenamefont {Asally}, \citenamefont {Ly},
  \citenamefont {Garcia-Ojalvo},\ and\ \citenamefont
  {S{\"u}el}}]{Prindle:2015hp}%
  \BibitemOpen
  \bibfield  {author} {\bibinfo {author} {\bibfnamefont {A}\ \bibnamefont
  {Prindle}}, \bibinfo {author} {\bibfnamefont {J}\ \bibnamefont {Liu}},
  \bibinfo {author} {\bibfnamefont {M}\ \bibnamefont {Asally}}, \bibinfo
  {author} {\bibfnamefont {S}\ \bibnamefont {Ly}}, \bibinfo {author}
  {\bibfnamefont {J}\ \bibnamefont {Garcia-Ojalvo}}, \ and\ \bibinfo
  {author} {\bibfnamefont {G~M}\ \bibnamefont {S{\"u}el}},\ }\bibfield
  {title} {\enquote {\bibinfo {title} {{Ion channels enable electrical
  communication in bacterial communities}},}\ }\href@noop {} {\bibfield
  {journal} {\bibinfo  {journal} {Nature}\ }\textbf {\bibinfo {volume} {527}},\
  \bibinfo {pages} {59--63} (\bibinfo {year} {2015})}\BibitemShut {NoStop}%
\bibitem [{\citenamefont {Liu}\ \emph {et~al.}(2015)\citenamefont {Liu},
  \citenamefont {Prindle}, \citenamefont {Humphries}, \citenamefont
  {Gabalda-Sagarra}, \citenamefont {Asally}, \citenamefont {Lee}, \citenamefont
  {Ly}, \citenamefont {Garcia-Ojalvo},\ and\ \citenamefont
  {S{\"u}el}}]{Liu:2015ht}%
  \BibitemOpen
  \bibfield  {author} {\bibinfo {author} {\bibfnamefont {J}\ \bibnamefont
  {Liu}}, \bibinfo {author} {\bibfnamefont {A}\ \bibnamefont {Prindle}},
  \bibinfo {author} {\bibfnamefont {J}\ \bibnamefont {Humphries}},
  \bibinfo {author} {\bibfnamefont {Mar{\c c}al}\ \bibnamefont
  {Gabalda-Sagarra}}, \bibinfo {author} {\bibfnamefont {M}\ \bibnamefont
  {Asally}}, \bibinfo {author} {\bibfnamefont {D~D}\ \bibnamefont
  {Lee}}, \bibinfo {author} {\bibfnamefont {S}\ \bibnamefont {Ly}}, \bibinfo
  {author} {\bibfnamefont {J}\ \bibnamefont {Garcia-Ojalvo}}, \ and\
  \bibinfo {author} {\bibfnamefont {G~M}\ \bibnamefont {S{\"u}el}},\
  }\bibfield  {title} {\enquote {\bibinfo {title} {{Metabolic co-dependence
  gives rise to collective oscillations within biofilms}},}\ }\href@noop {}
  {\bibfield  {journal} {\bibinfo  {journal} {Nature}\ }\textbf {\bibinfo
  {volume} {523}},\ \bibinfo {pages} {550--554} (\bibinfo {year}
  {2015})}\BibitemShut {NoStop}%
\bibitem [{\citenamefont {Humphries}\ \emph {et~al.}(2017)
\citenamefont  {Humphries}, \citenamefont {X}, \citenamefont {Liu}, \citenamefont
  {Prindle}, \citenamefont {Yuan}, \citenamefont {Arjes}, \citenamefont
  {Tsimring},\ and\ \citenamefont {S{\"u}el}}]{Humphries:2017gp}%
  \BibitemOpen
  \bibfield  {author} {\bibinfo {author} {\bibfnamefont {J}\
  \bibnamefont {Humphries}}, \bibinfo {author} {\bibfnamefont {L}\
  \bibnamefont {Xiong}}, \bibinfo {author} {\bibfnamefont {J}\
  \bibnamefont {Liu}}, \bibinfo {author} {\bibfnamefont {A}\ \bibnamefont
  {Prindle}}, \bibinfo {author} {\bibfnamefont {F}\ \bibnamefont {Yuan}},
  \bibinfo {author} {\bibfnamefont {H~A}\ \bibnamefont {Arjes}}, \bibinfo
  {author} {\bibfnamefont {L}\ \bibnamefont {Tsimring}}, \ and\ \bibinfo
  {author} {\bibfnamefont {G~M}\ \bibnamefont {S{\"u}el}},\ }\bibfield
  {title} {\enquote {\bibinfo {title} {{Species-Independent Attraction to
  Biofilms through Electrical Signaling}},}\ }\href@noop {} {\bibfield
  {journal} {\bibinfo  {journal} {Cell}\ }\textbf {\bibinfo {volume} {168}},\
  \bibinfo {pages} {200--209.e12} (\bibinfo {year} {2017})}\BibitemShut
  {NoStop}%
\bibitem [{\citenamefont {Kaplan}\ and\ \citenamefont
  {Greenberg}(1985)}]{Kaplan:1985ve}%
  \BibitemOpen
  \bibfield  {author} {\bibinfo {author} {\bibfnamefont {H~B}\ \bibnamefont
  {Kaplan}}\ and\ \bibinfo {author} {\bibfnamefont {E~P}\ \bibnamefont
  {Greenberg}},\ }\bibfield  {title} {\enquote {\bibinfo {title} {{Diffusion of
  autoinducer is involved in regulation of the Vibrio fischeri luminescence
  system.}}}\ }\href@noop {} {\bibfield  {journal} {\bibinfo  {journal}
  {Journal of Bacteriology}\ }\textbf {\bibinfo {volume} {163}},\ \bibinfo
  {pages} {1210--1214} (\bibinfo {year} {1985})}\BibitemShut {NoStop}%
\bibitem [{\citenamefont {Fuqua}\ \emph {et~al.}(1994)\citenamefont {Fuqua},
  \citenamefont {Winans},\ and\ \citenamefont {Greenberg}}]{Fuqua:1994wb}%
  \BibitemOpen
  \bibfield  {author} {\bibinfo {author} {\bibfnamefont {C~W}\
  \bibnamefont {Fuqua}}, \bibinfo {author} {\bibfnamefont {S~C}\
  \bibnamefont {Winans}}, \ and\ \bibinfo {author} {\bibfnamefont {E~P}\
  \bibnamefont {Greenberg}},\ }\bibfield  {title} {\enquote {\bibinfo {title}
  {{Quorum Sensing in Bacteria: the LuxR-LuxI family of cell density-responsive
  transcriptional regulators}},}\ }\href@noop {} {\bibfield  {journal}
  {\bibinfo  {journal} {J Bacteriology}\ }\textbf {\bibinfo {volume} {176}},\
  \bibinfo {pages} {269--275} (\bibinfo {year} {1994})}\BibitemShut {NoStop}%
\bibitem [{\citenamefont {Bassler}\ and\ \citenamefont
  {Losick}(2006)}]{Bassler:2006kd}%
  \BibitemOpen
  \bibfield  {author} {\bibinfo {author} {\bibfnamefont {B~L}\
  \bibnamefont {Bassler}}\ and\ \bibinfo {author} {\bibfnamefont {R}\
  \bibnamefont {Losick}},\ }\bibfield  {title} {\enquote {\bibinfo {title}
  {{Bacterially Speaking}},}\ }\href@noop {} {\bibfield  {journal} {\bibinfo
  {journal} {Cell}\ }\textbf {\bibinfo {volume} {125}},\ \bibinfo {pages}
  {237--246} (\bibinfo {year} {2006})}\BibitemShut {NoStop}%
\bibitem [{\citenamefont {Miller}\ and\ \citenamefont
  {Bassler}(2001)}]{Miller:2001uo}%
  \BibitemOpen
  \bibfield  {author} {\bibinfo {author} {\bibfnamefont {M~B}\ \bibnamefont
  {Miller}}\ and\ \bibinfo {author} {\bibfnamefont {B~L}\ \bibnamefont
  {Bassler}},\ }\bibfield  {title} {\enquote {\bibinfo {title} {{Quorum sensing
  in bacteria}},}\ }\href@noop {} {\bibfield  {journal} {\bibinfo  {journal}
  {Annual Reviews in Microbiology}\ }\textbf {\bibinfo {volume} {55}},\
  \bibinfo {pages} {165--199} (\bibinfo {year} {2001})}\BibitemShut {NoStop}%
\bibitem [{\citenamefont {Long}\ \emph {et~al.}(2009)\citenamefont {Long},
  \citenamefont {Tu}, \citenamefont {Wang}, \citenamefont {Mehta},
  \citenamefont {Ong}, \citenamefont {Bassler},\ and\ \citenamefont
  {Wingreen}}]{Long:2009bp}%
  \BibitemOpen
  \bibfield  {author} {\bibinfo {author} {\bibfnamefont {T}\ \bibnamefont
  {Long}}, \bibinfo {author} {\bibfnamefont {K~C}\ \bibnamefont {Tu}},
  \bibinfo {author} {\bibfnamefont {Y}\ \bibnamefont {Wang}}, \bibinfo
  {author} {\bibfnamefont {P}\ \bibnamefont {Mehta}}, \bibinfo {author}
  {\bibfnamefont {N~P}\ \bibnamefont {Ong}}, \bibinfo {author} {\bibfnamefont
  {B~L}\ \bibnamefont {Bassler}}, \ and\ \bibinfo {author} {\bibfnamefont
  {N~S}\ \bibnamefont {Wingreen}},\ }\bibfield  {title} {\enquote {\bibinfo
  {title} {{Quantifying the Integration of Quorum-Sensing Signals with
  Single-Cell Resolution}},}\ }\href@noop {} {\bibfield  {journal} {\bibinfo
  {journal} {PLoS Biol}\ }\textbf {\bibinfo {volume} {7}},\ \bibinfo {pages}
  {e1000068} (\bibinfo {year} {2009})}\BibitemShut {NoStop}%
\bibitem [{\citenamefont {P{\'e}rez-Vel{\'a}zquez}\ \emph
  {et~al.}(2016)\citenamefont {P{\'e}rez-Vel{\'a}zquez}, \citenamefont
  {G{\"o}lgeli},\ and\ \citenamefont
  {Garc{\'\i}a-Contreras}}]{PerezVelazquez:2016jm}%
  \BibitemOpen
  \bibfield  {author} {\bibinfo {author} {\bibfnamefont {J}\ \bibnamefont
  {P{\'e}rez-Vel{\'a}zquez}}, \bibinfo {author} {\bibfnamefont {M}\
  \bibnamefont {G{\"o}lgeli}}, \ and\ \bibinfo {author} {\bibfnamefont
  {R}\ \bibnamefont {Garc{\'\i}a-Contreras}},\ }\bibfield  {title}
  {\enquote {\bibinfo {title} {{s11538-016-0160-6}},}\ }\href@noop {}
  {\bibfield  {journal} {\bibinfo  {journal} {Bulletin of Mathematical
  Biology}\ }\textbf {\bibinfo {volume} {78}},\ \bibinfo {pages} {1585--1639}
  (\bibinfo {year} {2016})}\BibitemShut {NoStop}%
\bibitem [{\citenamefont {Basu}\ \emph {et~al.}(2005)\citenamefont {Basu},
  \citenamefont {Gerchman}, \citenamefont {Collins}, \citenamefont {Arnold},\
  and\ \citenamefont {Weiss}}]{Basu:2005cq}%
  \BibitemOpen
  \bibfield  {author} {\bibinfo {author} {\bibfnamefont {S}\ \bibnamefont
  {Basu}}, \bibinfo {author} {\bibfnamefont {Y}\ \bibnamefont {Gerchman}},
  \bibinfo {author} {\bibfnamefont {C~H}\ \bibnamefont {Collins}},
  \bibinfo {author} {\bibfnamefont {F~H}\ \bibnamefont {Arnold}}, \ and\
  \bibinfo {author} {\bibfnamefont {R}\ \bibnamefont {Weiss}},\ }\bibfield
  {title} {\enquote {\bibinfo {title} {{A synthetic multicellular system for
  programmed pattern formation}},}\ }\href@noop {} {\bibfield  {journal}
  {\bibinfo  {journal} {Nature}\ }\textbf {\bibinfo {volume} {434}},\ \bibinfo
  {pages} {1130--1134} (\bibinfo {year} {2005})}\BibitemShut {NoStop}%
\bibitem [{\citenamefont {Danino}\ \emph {et~al.}(2010)\citenamefont {Danino},
  \citenamefont {Mondrag{\'o}n-Palomino}, \citenamefont {Tsimring},\ and\
  \citenamefont {Hasty}}]{Danino:2010km}%
  \BibitemOpen
  \bibfield  {author} {\bibinfo {author} {\bibfnamefont {T}\ \bibnamefont
  {Danino}}, \bibinfo {author} {\bibfnamefont {O}\ \bibnamefont
  {Mondrag{\'o}n-Palomino}}, \bibinfo {author} {\bibfnamefont {L}\
  \bibnamefont {Tsimring}}, \ and\ \bibinfo {author} {\bibfnamefont {J}\
  \bibnamefont {Hasty}},\ }\bibfield  {title} {\enquote {\bibinfo {title} {{A
  synchronized quorum of genetic clocks}},}\ }\href@noop {} {\bibfield
  {journal} {\bibinfo  {journal} {Nature}\ }\textbf {\bibinfo {volume} {463}},\
  \bibinfo {pages} {326--330} (\bibinfo {year} {2010})}\BibitemShut {NoStop}%
\bibitem [{\citenamefont {Dilanji}\ \emph {et~al.}(2012)\citenamefont
  {Dilanji}, \citenamefont {Langebrake}, \citenamefont {De~Leenheer},\ and\
  \citenamefont {Hagen}}]{Dilanji:2012ds}%
  \BibitemOpen
  \bibfield  {author} {\bibinfo {author} {\bibfnamefont {G~E}\
  \bibnamefont {Dilanji}}, \bibinfo {author} {\bibfnamefont {J~B}\
  \bibnamefont {Langebrake}}, \bibinfo {author} {\bibfnamefont {P}\
  \bibnamefont {De~Leenheer}}, \ and\ \bibinfo {author} {\bibfnamefont
  {S~J}\ \bibnamefont {Hagen}},\ }\bibfield  {title} {\enquote {\bibinfo
  {title} {{Quorum Activation at a Distance: Spatiotemporal Patterns of Gene
  Regulation from Diffusion of an Autoinducer Signal}},}\ }\href@noop {}
  {\bibfield  {journal} {\bibinfo  {journal} {J. Am. Chem. Soc.}\ }\textbf
  {\bibinfo {volume} {134}},\ \bibinfo {pages} {5618--5626} (\bibinfo {year}
  {2012})}\BibitemShut {NoStop}%
\bibitem [{\citenamefont {Langebrake}\ \emph {et~al.}(2014)\citenamefont
  {Langebrake}, \citenamefont {Dilanji}, \citenamefont {Hagen},\ and\
  \citenamefont {De~Leenheer}}]{Langebrake:2014fg}%
  \BibitemOpen
  \bibfield  {author} {\bibinfo {author} {\bibfnamefont {J~B}\
  \bibnamefont {Langebrake}}, \bibinfo {author} {\bibfnamefont {G~E}\
  \bibnamefont {Dilanji}}, \bibinfo {author} {\bibfnamefont {S~J}\
  \bibnamefont {Hagen}}, \ and\ \bibinfo {author} {\bibfnamefont {P}\
  \bibnamefont {De~Leenheer}},\ }\bibfield  {title} {\enquote {\bibinfo {title}
  {{Traveling waves in response to a diffusing quorum sensing signal in
  spatially-extended bacterial colonies}},}\ }\href@noop {} {\bibfield
  {journal} {\bibinfo  {journal} {Journal of Theoretical Biology}\ }\textbf
  {\bibinfo {volume} {363}},\ \bibinfo {pages} {53--61} (\bibinfo {year}
  {2014})}\BibitemShut {NoStop}%
\bibitem [{\citenamefont {Ramalho}\ \emph {et~al.}(2016)\citenamefont
  {Ramalho}, \citenamefont {Meyer}, \citenamefont {M{\"u}ckl}, \citenamefont
  {Kapsner}, \citenamefont {Gerland},\ and\ \citenamefont
  {Simmel}}]{Ramalho:2016ki}%
  \BibitemOpen
  \bibfield  {author} {\bibinfo {author} {\bibfnamefont {T}\ \bibnamefont
  {Ramalho}}, \bibinfo {author} {\bibfnamefont {A}\ \bibnamefont {Meyer}},
  \bibinfo {author} {\bibfnamefont {A}\ \bibnamefont {M{\"u}ckl}},
  \bibinfo {author} {\bibfnamefont {K}\ \bibnamefont {Kapsner}},
  \bibinfo {author} {\bibfnamefont {U}\ \bibnamefont {Gerland}}, \ and\
  \bibinfo {author} {\bibfnamefont {F~C}\ \bibnamefont {Simmel}},\
  }\bibfield  {title} {\enquote {\bibinfo {title} {{Single Cell Analysis of a
  Bacterial Sender-Receiver System}},}\ }\href@noop {} {\bibfield  {journal}
  {\bibinfo  {journal} {PLoS ONE}\ }\textbf {\bibinfo {volume} {11}},\ \bibinfo
  {pages} {e0145829} (\bibinfo {year} {2016})}\BibitemShut {NoStop}%
\bibitem [{\citenamefont {Alberghini}\ \emph {et~al.}(2009)\citenamefont
  {Alberghini}, \citenamefont {Polone}, \citenamefont {Corich}, \citenamefont
  {Carlot}, \citenamefont {Seno}, \citenamefont {Trovato},\ and\ \citenamefont
  {Squartini}}]{Alberghini:2009gi}%
  \BibitemOpen
  \bibfield  {author} {\bibinfo {author} {\bibfnamefont {S}\ \bibnamefont
  {Alberghini}}, \bibinfo {author} {\bibfnamefont {E}\ \bibnamefont
  {Polone}}, \bibinfo {author} {\bibfnamefont {V}\ \bibnamefont
  {Corich}}, \bibinfo {author} {\bibfnamefont {M}\ \bibnamefont {Carlot}},
  \bibinfo {author} {\bibfnamefont {F}\ \bibnamefont {Seno}}, \bibinfo
  {author} {\bibfnamefont {A}\ \bibnamefont {Trovato}}, \ and\ \bibinfo
  {author} {\bibfnamefont {}\ \bibnamefont {Squartini}},\ }\bibfield
  {title} {\enquote {\bibinfo {title} {{Consequences of relative cellular
  positioning on quorum sensing and bacterial cell-to-cell communication}},}\
  }\href@noop {} {\bibfield  {journal} {\bibinfo  {journal} {FEMS Microbiology
  Letters}\ }\textbf {\bibinfo {volume} {292}},\ \bibinfo {pages} {149--161}
  (\bibinfo {year} {2009})}\BibitemShut {NoStop}%
\bibitem [{\citenamefont {Carslow}\ and\ \citenamefont
  {Jaeger}(1959)}]{Carslow:1959vx}%
  \BibitemOpen
  \bibfield  {author} {\bibinfo {author} {\bibfnamefont {H~S}\ \bibnamefont
  {Carslow}}\ and\ \bibinfo {author} {\bibfnamefont {J~C}\ \bibnamefont
  {Jaeger}},\ }\href@noop {} {\emph {\bibinfo {title} {{Conduction of heat in
  solids}}}}\ (\bibinfo  {publisher} {Oxford Clarendon Press},\ \bibinfo {year}
  {1959})\BibitemShut {NoStop}%
\bibitem [{\citenamefont {Kendall}(1948)}]{Kendall:1948df}%
  \BibitemOpen
  \bibfield  {author} {\bibinfo {author} {\bibfnamefont {D~G}\ \bibnamefont
  {Kendall}},\ }\bibfield  {title} {\enquote {\bibinfo {title} {{A form of wave
  propagation associated with the equation of heat conduction}},}\ }in\
  \href@noop {} {\emph {\bibinfo {booktitle} {Proceedings of the Cambridge
  Philosophical Society}}}\ (\bibinfo {year} {1948})\ pp.\ \bibinfo {pages}
  {591--594}\BibitemShut {NoStop}%
\bibitem [{\citenamefont {Fu}\ \emph {et~al.}(2012)\citenamefont {Fu},
  \citenamefont {Tang}, \citenamefont {Liu}, \citenamefont {Huang},
  \citenamefont {Hwa},\ and\ \citenamefont {Lenz}}]{Fu:2012en}%
  \BibitemOpen
  \bibfield  {author} {\bibinfo {author} {\bibfnamefont {X}\
  \bibnamefont {Fu}}, \bibinfo {author} {\bibfnamefont {L-H}\ \bibnamefont
  {Tang}}, \bibinfo {author} {\bibfnamefont {C}\ \bibnamefont {Liu}},
  \bibinfo {author} {\bibfnamefont {J-D}\ \bibnamefont {Huang}}, \bibinfo
  {author} {\bibfnamefont {T}\ \bibnamefont {Hwa}}, \ and\ \bibinfo
  {author} {\bibfnamefont {P}\ \bibnamefont {Lenz}},\ }\bibfield  {title}
  {\enquote {\bibinfo {title} {{Stripe Formation in Bacterial Systems with
  Density-Suppressed Motility}},}\ }\href@noop {} {\bibfield  {journal}
  {\bibinfo  {journal} {Phys Rev Lett}\ }\textbf {\bibinfo {volume} {108}},\
  \bibinfo {pages} {198102} (\bibinfo {year} {2012})}\BibitemShut {NoStop}%
\bibitem [{\citenamefont {Stewart}(2003)}]{Stewart:2003xy}%
  \BibitemOpen
  \bibfield  {author} {\bibinfo {author} {\bibfnamefont {P}~\bibnamefont
  {Stewart}},\ }\bibfield  {title} {\enquote {\bibinfo {title} {Diffusion in
  biofilms},}\ }\href@noop {} {\bibfield  {journal} {\bibinfo  {journal}
  {Journal of Bacteriology}\ }\textbf {\bibinfo {volume} {185}},\ \bibinfo
  {pages} {1485--1491} (\bibinfo {year} {2003})}\BibitemShut {NoStop}%
\bibitem [{\citenamefont {Charoenpanich}\ \emph {et~al.}(2013)\citenamefont
  {Charoenpanich}, \citenamefont {Meyer}, \citenamefont {Becker},\ and\
  \citenamefont {McIntosh}}]{Charoenpanich:2013gp}%
  \BibitemOpen
  \bibfield  {author} {\bibinfo {author} {\bibfnamefont {P}~\bibnamefont
  {Charoenpanich}}, \bibinfo {author} {\bibfnamefont {S}~\bibnamefont {Meyer}},
  \bibinfo {author} {\bibfnamefont {A}~\bibnamefont {Becker}}, \ and\ \bibinfo
  {author} {\bibfnamefont {M}~\bibnamefont {McIntosh}},\ }\bibfield  {title}
  {\enquote {\bibinfo {title} {{Temporal Expression Program of Quorum
  Sensing-Based Transcription Regulation in Sinorhizobium meliloti}},}\
  }\href@noop {} {\bibfield  {journal} {\bibinfo  {journal} {Journal of
  Bacteriology}\ }\textbf {\bibinfo {volume} {195}},\ \bibinfo {pages}
  {3224--3236} (\bibinfo {year} {2013})}\BibitemShut {NoStop}%
\bibitem [{\citenamefont {Krol}\ and\ \citenamefont
  {Becker}(2014)}]{Krol:2014jp}%
  \BibitemOpen
  \bibfield  {author} {\bibinfo {author} {\bibfnamefont {E}~\bibnamefont
  {Krol}}\ and\ \bibinfo {author} {\bibfnamefont {A}~\bibnamefont {Becker}},\
  }\bibfield  {title} {\enquote {\bibinfo {title} {{Rhizobial homologs of the
  fatty acid transporter FadL facilitate perception of long-chain
  acyl-homoserine lactone signals}},}\ }\href@noop {} {\bibfield  {journal}
  {\bibinfo  {journal} {Proc Nat Acad Sci USA}\ }\textbf {\bibinfo {volume}
  {111}},\ \bibinfo {pages} {10702--10707} (\bibinfo {year}
  {2014})}\BibitemShut {NoStop}%
\bibitem [{\citenamefont {Jones}\ \emph {et~al.}(2007)\citenamefont {Jones},
  \citenamefont {Kobayashi}, \citenamefont {Davies}, \citenamefont {Taga},\
  and\ \citenamefont {Walker}}]{Jones:2007dw}%
  \BibitemOpen
  \bibfield  {author} {\bibinfo {author} {\bibfnamefont {K~M}\
  \bibnamefont {Jones}}, \bibinfo {author} {\bibfnamefont {H}\
  \bibnamefont {Kobayashi}}, \bibinfo {author} {\bibfnamefont {B~W}\
  \bibnamefont {Davies}}, \bibinfo {author} {\bibfnamefont {M~E}\
  \bibnamefont {Taga}}, \ and\ \bibinfo {author} {\bibfnamefont {G~C}\
  \bibnamefont {Walker}},\ }\bibfield  {title} {\enquote {\bibinfo {title}
  {{How rhizobial symbionts invade plants: the Sinorhizobium--Medicago
  model}},}\ }\href@noop {} {\bibfield  {journal} {\bibinfo  {journal} {Nat Rev
  Micro}\ }\textbf {\bibinfo {volume} {5}},\ \bibinfo {pages} {619--633}
  (\bibinfo {year} {2007})}\BibitemShut {NoStop}%
\bibitem [{\citenamefont {Rinaudi}\ and\ \citenamefont
  {Giordano}(2010)}]{Rinaudi:2010ko}%
  \BibitemOpen
  \bibfield  {author} {\bibinfo {author} {\bibfnamefont {L~V}\
  \bibnamefont {Rinaudi}}\ and\ \bibinfo {author} {\bibfnamefont {W}\
  \bibnamefont {Giordano}},\ }\bibfield  {title} {\enquote {\bibinfo {title}
  {{An integrated view of biofilm formation in rhizobia}},}\ }\href@noop {}
  {\bibfield  {journal} {\bibinfo  {journal} {FEMS Microbiology Letters}\
  }\textbf {\bibinfo {volume} {304}},\ \bibinfo {pages} {1--11} (\bibinfo
  {year} {2010})}\BibitemShut {NoStop}%
\bibitem [{\citenamefont {Gurich}\ and\ \citenamefont
  {Gonzalez}(2009)}]{Gurich:2009hm}%
  \BibitemOpen
  \bibfield  {author} {\bibinfo {author} {\bibfnamefont {N}~\bibnamefont
  {Gurich}}\ and\ \bibinfo {author} {\bibfnamefont {J~E}\ \bibnamefont
  {Gonzalez}},\ }\bibfield  {title} {\enquote {\bibinfo {title} {{Role of
  Quorum Sensing in Sinorhizobium meliloti-Alfalfa Symbiosis}},}\ }\href@noop
  {} {\bibfield  {journal} {\bibinfo  {journal} {Journal of Bacteriology}\
  }\textbf {\bibinfo {volume} {191}},\ \bibinfo {pages} {4372--4382} (\bibinfo
  {year} {2009})}\BibitemShut {NoStop}%
\bibitem [{\citenamefont {McIntosh}\ \emph {et~al.}(2008)\citenamefont
  {McIntosh}, \citenamefont {Krol},\ and\ \citenamefont
  {Becker}}]{McIntosh:2008hc}%
  \BibitemOpen
  \bibfield  {author} {\bibinfo {author} {\bibfnamefont {M}~\bibnamefont
  {McIntosh}}, \bibinfo {author} {\bibfnamefont {E}~\bibnamefont {Krol}}, \
  and\ \bibinfo {author} {\bibfnamefont {A}~\bibnamefont {Becker}},\ }\bibfield
   {title} {\enquote {\bibinfo {title} {{Competitive and Cooperative Effects in
  Quorum-Sensing-Regulated Galactoglucan Biosynthesis in Sinorhizobium
  meliloti}},}\ }\href@noop {} {\bibfield  {journal} {\bibinfo  {journal}
  {Journal of Bacteriology}\ }\textbf {\bibinfo {volume} {190}},\ \bibinfo
  {pages} {5308--5317} (\bibinfo {year} {2008})}\BibitemShut {NoStop}%
\bibitem [{\citenamefont {Marketon}\ \emph {et~al.}(2002)\citenamefont
  {Marketon}, \citenamefont {Gronquist}, \citenamefont {Eberhard},\ and\
  \citenamefont {Gonzalez}}]{Marketon:2002ev}%
  \BibitemOpen
  \bibfield  {author} {\bibinfo {author} {\bibfnamefont {M~M}\ \bibnamefont
  {Marketon}}, \bibinfo {author} {\bibfnamefont {M~R}\ \bibnamefont
  {Gronquist}}, \bibinfo {author} {\bibfnamefont {A}~\bibnamefont {Eberhard}},
  \ and\ \bibinfo {author} {\bibfnamefont {J~E}\ \bibnamefont {Gonzalez}},\
  }\bibfield  {title} {\enquote {\bibinfo {title} {{Characterization of the
  Sinorhizobium meliloti sinR/sinI Locus and the Production of Novel N-Acyl
  Homoserine Lactones}},}\ }\href@noop {} {\bibfield  {journal} {\bibinfo
  {journal} {Journal of Bacteriology}\ }\textbf {\bibinfo {volume} {184}},\
  \bibinfo {pages} {5686--5695} (\bibinfo {year} {2002})}\BibitemShut {NoStop}%
\bibitem [{\citenamefont {Gao}\ \emph {et~al.}(2005)\citenamefont {Gao},
  \citenamefont {Chen}, \citenamefont {Eberhard}, \citenamefont {Gronquist},
  \citenamefont {Robinson}, \citenamefont {Rolfe},\ and\ \citenamefont
  {Bauer}}]{Gao:2005kp}%
  \BibitemOpen
  \bibfield  {author} {\bibinfo {author} {\bibfnamefont {M}~\bibnamefont
  {Gao}}, \bibinfo {author} {\bibfnamefont {H}~\bibnamefont {Chen}}, \bibinfo
  {author} {\bibfnamefont {A}~\bibnamefont {Eberhard}}, \bibinfo {author}
  {\bibfnamefont {M~R}\ \bibnamefont {Gronquist}}, \bibinfo {author}
  {\bibfnamefont {J~B}\ \bibnamefont {Robinson}}, \bibinfo {author}
  {\bibfnamefont {B~G}\ \bibnamefont {Rolfe}}, \ and\ \bibinfo {author}
  {\bibfnamefont {W~D}\ \bibnamefont {Bauer}},\ }\bibfield  {title} {\enquote
  {\bibinfo {title} {{sinI- and expR-Dependent Quorum Sensing in Sinorhizobium
  meliloti}},}\ }\href@noop {} {\bibfield  {journal} {\bibinfo  {journal}
  {Journal of Bacteriology}\ }\textbf {\bibinfo {volume} {187}},\ \bibinfo
  {pages} {7931--7944} (\bibinfo {year} {2005})}\BibitemShut {NoStop}%
\bibitem [{\citenamefont {McIntosh}\ \emph {et~al.}(2009)\citenamefont
  {McIntosh}, \citenamefont {Meyer},\ and\ \citenamefont
  {Becker}}]{McIntosh:2009db}%
  \BibitemOpen
  \bibfield  {author} {\bibinfo {author} {\bibfnamefont {M}\ \bibnamefont
  {McIntosh}}, \bibinfo {author} {\bibfnamefont {Stefan}\ \bibnamefont
  {Meyer}}, \ and\ \bibinfo {author} {\bibfnamefont {A}\ \bibnamefont
  {Becker}},\ }\bibfield  {title} {\enquote {\bibinfo {title} {{Novel
  Sinorhizobium melilotiquorum sensing positive and negative regulatory
  feedback mechanisms respond to phosphate availability}},}\ }\href@noop {}
  {\bibfield  {journal} {\bibinfo  {journal} {Molecular Microbiology}\ }\textbf
  {\bibinfo {volume} {74}},\ \bibinfo {pages} {1238--1256} (\bibinfo {year}
  {2009})}\BibitemShut {NoStop}%
\bibitem [{\citenamefont {Schl{\"u}ter}\ \emph {et~al.}(2015)\citenamefont
  {Schl{\"u}ter}, \citenamefont {Czuppon}, \citenamefont {Schauer},
  \citenamefont {Pfaffelhuber}, \citenamefont {McIntosh},\ and\ \citenamefont
  {Becker}}]{Schluter:2015ax}%
  \BibitemOpen
  \bibfield  {author} {\bibinfo {author} {\bibfnamefont {J P}\
  \bibnamefont {Schl{\"u}ter}}, \bibinfo {author} {\bibfnamefont {P}\
  \bibnamefont {Czuppon}}, \bibinfo {author} {\bibfnamefont {O}\
  \bibnamefont {Schauer}}, \bibinfo {author} {\bibfnamefont {P}\
  \bibnamefont {Pfaffelhuber}}, \bibinfo {author} {\bibfnamefont {M}\
  \bibnamefont {McIntosh}}, \ and\ \bibinfo {author} {\bibfnamefont {A}\
  \bibnamefont {Becker}},\ }\bibfield  {title} {\enquote {\bibinfo {title}
  {Classification of phenotypic subpopulations in isogenic bacterial cultures
  by triple promoter probing at single cell level},}\ }\href {\doibase
  https://doi.org/10.1016/j.jbiotec.2015.01.021} {\bibfield  {journal}
  {\bibinfo  {journal} {Journal of Biotechnology}\ }\textbf {\bibinfo {volume}
  {198}},\ \bibinfo {pages} {3 -- 14} (\bibinfo {year} {2015})}\BibitemShut
  {NoStop}%
\bibitem [{\citenamefont {Ortiz}\ and\ \citenamefont
  {Endy}(2012)}]{Ortiz:2012gq}%
  \BibitemOpen
  \bibfield  {author} {\bibinfo {author} {\bibfnamefont {M~E}\
  \bibnamefont {Ortiz}}\ and\ \bibinfo {author} {\bibfnamefont {D}\
  \bibnamefont {Endy}},\ }\bibfield  {title} {\enquote {\bibinfo {title}
  {{Engineered cell-cell communication via DNA messaging.}}}\ }\href@noop {}
  {\bibfield  {journal} {\bibinfo  {journal} {J Biol Eng}\ }\textbf {\bibinfo
  {volume} {6}},\ \bibinfo {pages} {1--10} (\bibinfo {year}
  {2012})}\BibitemShut {NoStop}%
\bibitem [{\citenamefont {Dohlemann}\ \emph {et~al.}(2017)\citenamefont
  {Dohlemann}, \citenamefont {Wagner}, \citenamefont {Happel}, \citenamefont
  {Carrillo}, \citenamefont {Sobetzko}, \citenamefont {Erb}, \citenamefont
  {Thanbichler},\ and\ \citenamefont {Becker}}]{Dohlemann:2017xy}%
  \BibitemOpen
  \bibfield  {author} {\bibinfo {author} {\bibfnamefont {J}\
  \bibnamefont {Dohlemann}}, \bibinfo {author} {\bibfnamefont {M}\
  \bibnamefont {Wagner}}, \bibinfo {author} {\bibfnamefont {C}\
  \bibnamefont {Happel}}, \bibinfo {author} {\bibfnamefont {MM}\
  \bibnamefont {Carrillo}}, \bibinfo {author} {\bibfnamefont {P}\
  \bibnamefont {Sobetzko}}, \bibinfo {author} {\bibfnamefont {T~J.}\
  \bibnamefont {Erb}}, \bibinfo {author} {\bibfnamefont {M}\ \bibnamefont
  {Thanbichler}}, \ and\ \bibinfo {author} {\bibfnamefont {A}\ \bibnamefont
  {Becker}},\ }\bibfield  {title} {\enquote {\bibinfo {title} {A family of
  single copy repabc-type shuttle vectors stably maintained in the
  alpha-proteobacterium sinorhizobium meliloti},}\ }\href {\doibase
  10.1021/acssynbio.6b00320} {\bibfield  {journal} {\bibinfo  {journal} {ACS
  Synthetic Biology}\ }\textbf {\bibinfo {volume} {6}},\ \bibinfo {pages}
  {968--984} (\bibinfo {year} {2017})},\ \bibinfo {note} {pMID:
  28264559}\BibitemShut {NoStop}%
\bibitem [{\citenamefont {Nagai}\ \emph {et~al.}(2002)\citenamefont {Nagai},
  \citenamefont {Ibata}, \citenamefont {Park}, \citenamefont {Kubota},
  \citenamefont {Mikoshiba},\ and\ \citenamefont {Miyawaki}}]{Nagai:2002wb}%
  \BibitemOpen
  \bibfield  {author} {\bibinfo {author} {\bibfnamefont {T}~\bibnamefont
  {Nagai}}, \bibinfo {author} {\bibfnamefont {K}~\bibnamefont {Ibata}},
  \bibinfo {author} {\bibfnamefont {E~S}\ \bibnamefont {Park}}, \bibinfo
  {author} {\bibfnamefont {M}\ \bibnamefont {Kubota}}, \bibinfo {author}
  {\bibfnamefont {K}\ \bibnamefont {Mikoshiba}}, \ and\ \bibinfo
  {author} {\bibfnamefont {A}\ \bibnamefont {Miyawaki}},\ }\bibfield
  {title} {\enquote {\bibinfo {title} {{A variant of yellow fluorescent protein
  with fast and efficient maturation for cell-biological applications}},}\
  }\href@noop {} {\bibfield  {journal} {\bibinfo  {journal} {Nature
  Biotechnology}\ }\textbf {\bibinfo {volume} {20}},\ \bibinfo {pages} {87--90}
  (\bibinfo {year} {2002})}\BibitemShut {NoStop}%
\bibitem [{\citenamefont {Sch{\"a}fer}\ \emph {et~al.}(1994)\citenamefont
  {Sch{\"a}fer}, \citenamefont {Tauch}, \citenamefont {J{\"a}ger},
  \citenamefont {Kalinowski}, \citenamefont {Thierbach},\ and\ \citenamefont
  {P{\"u}hler}}]{Schaefer:1994ax}%
  \BibitemOpen
  \bibfield  {author} {\bibinfo {author} {\bibfnamefont {A}\ \bibnamefont
  {Sch{\"a}fer}}, \bibinfo {author} {\bibfnamefont {A}\ \bibnamefont
  {Tauch}}, \bibinfo {author} {\bibfnamefont {W}\ \bibnamefont
  {J{\"a}ger}}, \bibinfo {author} {\bibfnamefont {J}\ \bibnamefont
  {Kalinowski}}, \bibinfo {author} {\bibfnamefont {G}\ \bibnamefont
  {Thierbach}}, \ and\ \bibinfo {author} {\bibfnamefont {A}\ \bibnamefont
  {P{\"u}hler}},\ }\bibfield  {title} {\enquote {\bibinfo {title} {Small
  mobilizable multi-purpose cloning vectors derived from the escherichia coli
  plasmids pk18 and pk19: selection of defined deletions in the chromosome of
  corynebacterium glutamicum},}\ }\href {\doibase
  https://doi.org/10.1016/0378-1119(94)90324-7} {\bibfield  {journal} {\bibinfo
   {journal} {Gene}\ }\textbf {\bibinfo {volume} {145}},\ \bibinfo {pages} {69
  -- 73} (\bibinfo {year} {1994})}\BibitemShut {NoStop}%
\bibitem [{\citenamefont {Bahlawane}\ \emph {et~al.}(2008)\citenamefont
  {Bahlawane}, \citenamefont {McIntosh}, \citenamefont {Krol},\ and\
  \citenamefont {Becker}}]{Bahlawane:2008dg}%
  \BibitemOpen
  \bibfield  {author} {\bibinfo {author} {\bibfnamefont {C}\
  \bibnamefont {Bahlawane}}, \bibinfo {author} {\bibfnamefont {M}\
  \bibnamefont {McIntosh}}, \bibinfo {author} {\bibfnamefont {E}\
  \bibnamefont {Krol}}, \ and\ \bibinfo {author} {\bibfnamefont {A}\
  \bibnamefont {Becker}},\ }\bibfield  {title} {\enquote {\bibinfo {title}
  {{Sinorhizobium melilotiRegulator MucR Couples Exopolysaccharide Synthesis
  and Motility}},}\ }\href@noop {} {\bibfield  {journal} {\bibinfo  {journal}
  {MPMI}\ }\textbf {\bibinfo {volume} {21}},\ \bibinfo {pages} {1498--1509}
  (\bibinfo {year} {2008})}\BibitemShut {NoStop}%
\bibitem [{\citenamefont {Shaner}\ \emph {et~al.}(2004)\citenamefont {Shaner},
  \citenamefont {Campbell}, \citenamefont {Steinbach}, \citenamefont
  {Giepmans}, \citenamefont {Palmer},\ and\ \citenamefont
  {Tsien}}]{Shaner:2004ax}%
  \BibitemOpen
  \bibfield  {author} {\bibinfo {author} {\bibfnamefont {N~C}\
  \bibnamefont {Shaner}}, \bibinfo {author} {\bibfnamefont {R~E.}\
  \bibnamefont {Campbell}}, \bibinfo {author} {\bibfnamefont {P~A}\
  \bibnamefont {Steinbach}}, \bibinfo {author} {\bibfnamefont {B N~G}\
  \bibnamefont {Giepmans}}, \bibinfo {author} {\bibfnamefont {A~E}\
  \bibnamefont {Palmer}}, \ and\ \bibinfo {author} {\bibfnamefont {R}\
  \bibnamefont {Tsien}},\ }\bibfield  {title} {\enquote {\bibinfo {title}
  {Improved monomeric red, orange and yellow fluorescent proteins derived from
  discosoma sp. red fluorescent protein},}\ }\href@noop {} {\bibfield
  {journal} {\bibinfo  {journal} {Nature Biotechnology}\ }\textbf {\bibinfo
  {volume} {22}},\ \bibinfo {pages} {1567--1572} (\bibinfo {year}
  {2004})}\BibitemShut {NoStop}%
\bibitem [{\citenamefont {H{\"u}bner}\ \emph {et~al.}(1991)\citenamefont
  {H{\"u}bner}, \citenamefont {Willison}, \citenamefont {Vignais},\ and\
  \citenamefont {Bickle}}]{Hubner:1991uy}%
  \BibitemOpen
  \bibfield  {author} {\bibinfo {author} {\bibfnamefont {P}~\bibnamefont
  {H{\"u}bner}}, \bibinfo {author} {\bibfnamefont {J~C}\ \bibnamefont
  {Willison}}, \bibinfo {author} {\bibfnamefont {P~M}\ \bibnamefont {Vignais}},
  \ and\ \bibinfo {author} {\bibfnamefont {T~A}\ \bibnamefont {Bickle}},\
  }\bibfield  {title} {\enquote {\bibinfo {title} {{Expression of regulatory
  nif genes in Rhodobacter capsulatus}},}\ }\href@noop {} {\bibfield  {journal}
  {\bibinfo  {journal} {J Bacteriology}\ }\textbf {\bibinfo {volume} {173}},\
  \bibinfo {pages} {2993--2999} (\bibinfo {year} {1991})}\BibitemShut {NoStop}%
\bibitem [{\citenamefont {Rizzo}\ \emph {et~al.}(2004)\citenamefont {Rizzo},
  \citenamefont {Springer}, \citenamefont {Granada},\ and\ \citenamefont
  {Piston}}]{Rizzo:2004hz}%
  \BibitemOpen
  \bibfield  {author} {\bibinfo {author} {\bibfnamefont {M~A}\ \bibnamefont
  {Rizzo}}, \bibinfo {author} {\bibfnamefont {G~H}\ \bibnamefont
  {Springer}}, \bibinfo {author} {\bibfnamefont {B}\ \bibnamefont
  {Granada}}, \ and\ \bibinfo {author} {\bibfnamefont {D~W}\ \bibnamefont
  {Piston}},\ }\bibfield  {title} {\enquote {\bibinfo {title} {{An improved
  cyan fluorescent protein variant useful for FRET}},}\ }\href@noop {}
  {\bibfield  {journal} {\bibinfo  {journal} {Nature Biotechnology}\ }\textbf
  {\bibinfo {volume} {22}},\ \bibinfo {pages} {445--449} (\bibinfo {year}
  {2004})}\BibitemShut {NoStop}%
\bibitem [{\citenamefont {Zhan}\ \emph {et~al.}(1991)\citenamefont {Zhan},
  \citenamefont {Lee},\ and\ \citenamefont {Leigh}}]{Zhan:1991te}%
  \BibitemOpen
  \bibfield  {author} {\bibinfo {author} {\bibfnamefont {H}~\bibnamefont
  {Zhan}}, \bibinfo {author} {\bibfnamefont {C~C}\ \bibnamefont {Lee}}, \ and\
  \bibinfo {author} {\bibfnamefont {J~A}\ \bibnamefont {Leigh}},\ }\bibfield
  {title} {\enquote {\bibinfo {title} {{Induction of the second
  exopolysaccharide (EPSb) in Rhizobium meliloti SU47 by low phosphate
  concentrations}},}\ }\href@noop {} {\bibfield  {journal} {\bibinfo  {journal}
  {J Bacteriology}\ }\textbf {\bibinfo {volume} {173}},\ \bibinfo {pages}
  {7391--7394} (\bibinfo {year} {1991})}\BibitemShut {NoStop}%
\end{thebibliography}

%

\clearpage
\end{document}